\documentclass[fleqn,usenatbib]{mnras}

\usepackage{newtxtext,newtxmath}

\usepackage[T1]{fontenc}
\usepackage{ae,aecompl}

\usepackage{amsmath}	
\usepackage{amssymb}	
\usepackage{graphicx}	
\usepackage{amsmath}	
\usepackage{amssymb}	
\usepackage{multicol}        
\usepackage{bm}		
\usepackage{pdflscape}	
\usepackage[]{units}
\usepackage{hyperref}

\newcommand{\be}{\begin{equation}}
\newcommand{\ee}{\end{equation}}

\newcommand{\etal}{et al.}

\newcommand{\msun}{M_{\sun}}

\newcommand{\FIREurl}{\url{http://fire.northwestern.edu}}
\newcommand{\gizmourl}{\url{www.tapir.caltech.edu/~phopkins/Site/GIZMO.html}}

\newcommand{\acknowledgments}{\begin{small}\section*{Acknowledgments}\end{small}}
\newcommand\altaffilmark[1]{$^{#1}$}
\newcommand\altaffiltext[1]{$^{#1}$}
\title[When Feedback Fails]{When Feedback Fails: The Scaling and Saturation of Star Formation Efficiency}
\vspace{-0.5cm}

\vspace{-0.2cm}
\author[Grudi{\'c} \etal]{
\parbox[t]{\textwidth}{ 
Michael Y. Grudi{\'c}\thanks{E-mail: mgrudich@caltech.edu}\altaffilmark{1},
Philip F. Hopkins\altaffilmark{1}, Claude-Andr\'{e} Faucher-Gigu\`{e}re\altaffilmark{2}, Eliot Quataert\altaffilmark{3}, Norman Murray \altaffilmark{4,5}, Du\v{s}an Kere\v{s}\altaffilmark{6}
} 
\vspace*{6pt} \\
\altaffiltext{1}{TAPIR, Mailcode 350-17, California Institute of Technology, Pasadena, CA 91125, USA} \\
\altaffiltext{2}{Department of Physics and Astronomy and CIERA, Northwestern University, 2145 Sheridan Road, Evanston, IL 60208, USA} \\
\altaffiltext{3}{Department of Astronomy and Theoretical Astrophysics Center, University of California Berkeley, Berkeley, CA 94720} \\
\altaffiltext{4}{Canadian Institute for Theoretical Astrophysics, 
60 St.\ George Street, University of Toronto, ON M5S 3H8, Canada} \\
\altaffiltext{5}{Canada Research Chair in Astrophysics} \\
\altaffiltext{6}{Department of Physics, Center for Astrophysics and Space Science, UC San Diego, 9500 Gilman Drive, La Jolla, CA 92093} \\ 
\vspace{-0.5cm}
}
\date{Submitted to MNRAS, December, 2016\vspace{-0.6cm}}

\hypersetup{final}
\begin{document}
\maketitle
\label{firstpage}

\vspace{-0.2cm}
\begin{abstract}
We present a suite of 3D multi-physics MHD simulations following star formation in isolated turbulent molecular gas disks ranging from 5 to 500 parsecs in radius. These simulations are designed to survey the range of surface densities between those typical of Milky Way GMCs ($\sim \unit[10^2]{\msun pc^{-2}}$) and extreme ULIRG environments ($\sim \unit[10^4]{\msun pc^{-2}}$) so as to map out the scaling of the cloud-scale star formation efficiency (SFE) between these two regimes. The simulations include prescriptions for supernova, stellar wind, and radiative feedback, which we find to be essential in determining both the instantaneous per-freefall ($\epsilon_{ff}$) and integrated ($\epsilon_{int}$) star formation efficiencies. In all simulations, the gas disks form stars until a critical stellar surface density has been reached and the remaining gas is blown out by stellar feedback. We find that surface density is a good predictor of $\epsilon_{int}$, as suggested by analytic force balance arguments from previous works. SFE eventually saturates to $\sim 1$ at high surface density. We also find a proportional relationship between $\epsilon_{ff}$ and $\epsilon_{int}$, implying that star formation is feedback-moderated even over very short time-scales in isolated clouds. These results have implications for star formation in galactic disks, the nature and fate of nuclear starbursts, and the formation of bound star clusters. The scaling of $\epsilon_{ff}$ with surface density is not consistent with the notion that $\epsilon_{ff}$ is always $\sim 1\%$ on the scale of GMCs, but our predictions recover the $\sim 1\%$ value for GMC parameters similar to those found in sprial galaxies, including our own.
\end{abstract}

\begin{keywords}
galaxies: star formation --- galaxies: starburst --- galaxies: active --- galaxies: nuclei --- 
galaxies: star clusters: general \vspace{-0.5cm}
\end{keywords}

\section{Introduction}\label{intro}
Typically, star formation in the observed Universe is inefficient in any sense of the word. Star formation is observed to occur in giant molecular clouds (GMCs) formed in galactic disks, and the per-freefall star formation efficiency of a star-forming region may be parametrized as:
\begin{equation}
\dot{M}_\star \left(t\right)= \epsilon_{ff}\left(t\right) \frac{M_{gas}\left(t\right)}{t_{ff}\left(t\right)},
\label{eq:eff}
\end{equation}
where $\dot{M}_\star$ is the star formation rate, $M_{gas}$ is the gas mass ``available'' to form stars (observationally, the mass of molecular or dengas as obtained from a tracer such as CO or HCN), and $t_{ff}\left(t\right)$ is the local gravitational freefall time. $\epsilon_{ff}$ is the fraction of available gas converted to stars per $t_{ff}$; on galactic ($\sim \unit[]{kpc}$) scales, $\epsilon_{ff}$ has been estimated by fitting to the relation:
\vspace{-0.05in}
\begin{equation}
\Sigma_{SFR} = \epsilon_{ff}^{gal} \Sigma_{gas} t_{ff}^{-1},
\label{KS}
\end{equation}
where $\Sigma_{SFR}$ is the projected density of star formation in the disk, $\Sigma_{gas}$ is the projected (cold) gas density, $t_{ff}$ is the local freefall time evaluated from the galaxy's scale height-averaged density, and $\epsilon_{ff}^{gal}$ has been found to be $\sim 0.02$  \citep{kennicutt98}. Thus, a typical galaxy converts only $2\%$ of its potentially star-forming gas into stars each freefall time, despite the tendency of self-gravitating cold gas clouds to fragment and contract nearly all of their gas mass to high densities within only a few $t_{ff}$. Clearly, some physical mechanism is responsible for the moderation of star formation.

Recently, the FIRE\footnote{\FIREurl} (Feedback In Realistic Environments) simulations \citep{hopkins:2013.fire,fire2} have demonstrated that the inefficiency of star formation in galaxies formed within the $\mathrm{\Lambda}$CDM cosmology can be explained by stellar feedback, obtaining good agreement with \citet{kennicutt98} independent of the numerical resolution-scale star formation model. As stars form in dense GMCs within a galaxy, some combination of of photoionization heating, radiation pressure, stellar winds, and possibly supernovae blow out the remaining gas in the cloud, terminating star formation locally. The young stars formed inject momentum, mass, and energy into the surrounding ISM, which prevents the runaway vertical collapse of the galactic disk by providing turbulent support, and the rates of turbulent dissipation and momentum injection are in equilibrium when $\epsilon_{ff}^{gal} \sim 0.02$  (see \citet{thompson:rad.pressure, ostriker.shetty:2011,cafg:sf.fb.reg.kslaw, orr:2016.what.fires.up.SF}).

However, this mechanism only explains the rate of star formation on galactic scales: $\epsilon_{ff}^{gal}$ emerges from an established equilibrium over the formation and disruption of many GMCs, and is distinct from the value of $\epsilon_{ff}$ for a single GMC. Since star formation in a GMC must cease once it is disrupted, there exists another quantity of interest in characterizing the efficiency of star formation, the integrated SFE:
\begin{equation}
\epsilon_{int} = \frac{M_\star}{M_{tot}},
\label{def:sfe}
\end{equation}
where $M_\star$ is the final mass of stars formed and $M_{tot}$ is the mass of the initial gas cloud. In Milky Way GMCs, the median value of $\epsilon_{int}$ is on the order of $1\%$, \citep{evans:2009.sfe, murray:2010.sfe.mw.gmc,lee:2016.gmc.eff,vuti:2016.gmcs} with a large observed scatter of $\unit[0.8]{dex}$ \citep{murray:2010.sfe.mw.gmc,lee:2016.gmc.eff}. However, there is evidence that $\epsilon_{int}$ is much higher in denser conditions: \citet{murray:molcloud.disrupt.by.rad.pressure} points out that the masses of GMCs \citep{keto:2005.m82.gmcs} and young star clusters \citep{mccrady:m82.sscs} in the M82 starburst galaxy are of a similar mass scale, suggesting that $\epsilon_{int}$ is of order unity at the greater surface densities of such regions. Indeed, the existence of young, bound star clusters {\it in general} may physically require high integrated SFE on at least some local scale \citep{tutukov:1978,hills:1980,elmegreen:1983,mathieu:1983,elmegreen:1997.open.closed.cluster.same.mf.form}. Recent observations of young massive clusters (YMCs) have also suggested a time constraint of $<\unit[4]{Myr}$ for cluster formation within the disk of M83 \citep{hollyhead:2015.m83.ymcs}, only twice the typical GMC freefall time in the central region of M83 \citep{freeman:2017.m83.gmcs}, suggesting that cluster formation may also be a dynamically-fast process. Therefore, it is necessary to explore ways in which the efficiency of star formation, both in terms of $\epsilon_{ff}$ and $\epsilon_{int}$, can scale from Milky Way-like values of $\sim 1\%$ to greater values. Since stellar feedback is responsible for the eventual disruption of molecular clouds against gravity, it is likely that the balance of these two forces plays a major role in determining both the speed and integrated efficiency of star formation at sub-$\unit[]{kpc}$ scales. 

In this paper, we focus on the detailed behaviour of a single star formation episode at high resolution: we present 3D MHD simulations of star-forming gas disks which use the numerical treatments of cooling, star formation and stellar feedback of \citet{fire2} to answer certain basic questions about star formation in local galactic environments:
\begin{itemize}
\item Given an initial self-gravitating gas distribution, what is the resulting star formation history? In particular, what determines the observable quantities $\epsilon_{ff}$ and $\epsilon_{int}$, and how are they related?
\item How do the initial parameters of the gas cloud map onto the properties of the formed stellar system?
\item Which physical mechanisms have the greatest effect upon the answers to these questions?
\end{itemize}

The general approach of this study is to suppose some generic initial conditions for an isolated gas disk, neglecting its interaction with the surrounding galactic environment. This approximation makes sense for simulations spanning no more than a few dynamical times (which we shall show to be the case) and allows us to achieve relatively high spatial and mass resolution in the region of interest for modest computational cost.

This physics problem is most conventionally applicable to star-forming GMCs, but really any region in which the dynamical time is not significantly longer than the main sequence lifetime of massive stars ($\sim \unit[3]{Myr}$) should be unstable to runaway star formation and the eventual blowout of the gas component \citep{torrey:2016.feedback.instability}. The central regions of ultraluminous infrared galaxies (ULIRGs) may have large gas fractions and short dynamical times \citep{downes.solomon:ulirgs,bryant.scoville:ulirgs.co}, so for the purposes of our problem they may effectively behave as one super-GMC with particularly high ($>\unit[10^3]{\msun\,pc^{-2}}$) surface density. Our simulations, which probe these surface densities, can therefore also serve as models of gas-rich nuclear disks, which host the most extreme star formation events in the local Universe.

This paper is structured as follows: in Section \ref{sec:derivation}, we describe a simple model of a gas-rich, star-forming disk, and predict its general behaviour from the physical arguments. In Section \ref{sec:sims}, we describe the methods for our simulations, their initial conditions, and the scope of our survey of physics and simulation parameters. In Section \ref{sec:results} we present the results of the simulations concerning the global properties of the star-forming clouds: the overall behaviour of the simulated clouds, the isolated effects of various physical mechanisms, the per-freefall ($\epsilon_{ff}$) and integrated ($\epsilon_{int}$) star formation efficiency. Finally, in Section \ref{sec:discussion} we discuss some applications, implications and limitations of our results and outline future studies on the more detailed aspects of the mode of star formation we have simulated.

\section{A Star-forming Disk Model}
\label{sec:derivation}

To guide the methodology of the numerical study, we first review some basic theory of star formation and construct a simple model that captures the essential physics of how feedback determines the SFE of a gas-rich star-forming disk over short dynamical timescales. Consider an initially-uniform disk of mass $M$, radius $R$, and scale height $h$ that initially consists of only gas. Averaged over the diameter of the disk, the initial surface density is then:
\begin{equation}
\Sigma_{tot,0} = \Sigma_{gas}(t=0) = \frac{M}{\pi R^2}.
\end{equation}
\subsection{Time-scales for star formation}
The longest possible time-scale for gravitational collapse within the model disk is the freefall time $t_{ff,0}$ derived from the system's physical parameters $M$ and $R$:
\begin{equation}
t_{ff,0} = \frac{\pi}{2} \sqrt{\frac{R^3}{2 G M}} = \unit[2]{Myr}\left(\frac{R}{\unit[50]{pc}}\right)^{\frac{1}{2}}\left(\frac{\Sigma_{tot,0}}{\unit[10^3]{\msun\,pc^{-2}}}\right)^{-\frac{1}{2}},
\label{tff0}
\end{equation}
which is proportional to the outer orbital period of the disk. This is the longest relevant time-scale in the problem, since we neglect environmental interactions. $t_{ff,0}$ may overestimate the typical gravitational collapse time of a typical gas parcel, as we expect that if star formation is to occur then the dynamics are driving mass to greater-than-average densities with correspondingly shorter freefall times. Specifically, isothermal, self-gravitating turbulence has been found to produce a density PDF with a high-density power-law tail due to gravity \citep{kritsuk:2011.density.pdf.power.law}, and at lower densities a log-normal form, as emerges in isothermal turbulence without gravity \citep{vazquez-semadeni:1994.turb.density.pdf, padoan:1997.density.pdf, padoan:1999.density.pdf}. The only characteristic density is the peak of this distribution, so we define a shorter freefall time in terms of the median gas density $\rho_{50}$ (equivalently, number density $n_{50}$)\footnote{Note that we use the median, and not the mass-weighted mean gas density used for determining $t_{ff}$ in \citet{krumholz:2011.rhd.starcluster.sim} and \citet{2014MNRAS.439.3420M}. The mass-weighted mean is less suitable for estimating $t_{ff}$ in the middle of star formation because the high-density power-law tail in the density PDF biases it toward high densities. We also find that it is not robust with respect to simulation resolution, as higher resolutions will resolve more of the power-law tail. The median density generally lies near the peak of the density PDF, and is robust with respect to resolution.}: 
\begin{equation}
t_{ff,50} = \sqrt{\frac{3\pi}{32 G \rho_{50}}} = \unit[1.6]{Myr} \left(\frac{n_{50}}{10^3 cm^{-3}}\right)^{-\frac{1}{2}},
\label{def:tffmode}
\end{equation}

where $n_{50}$ is the median particle number density. $t_{ff,50}$ will generally be a more reasonable unit for the gas depletion time, and hence for comparing values of $\epsilon_{ff}$.

In the parameter space relevant to star formation in the local Universe, the cooling time of gas that is metal-enriched or molecular is generally much less than both $t_{ff,0}$ and $t_{ff,50}$. Therefore, in absence of stars or external inputs, any thermal energy supporting against self-gravity will quickly radiate away. If the disk has some initial turbulent velocity dispersion, that energy too will be cooled away by shocks over $\sim t_{ff,50}$. Without some imposed stabilizing force the disk will be subject to gravitational instability, fragmentation, and star formation. 

The process of fragmentation involves a runaway collapse to protostellar densities. If an initially-smooth disk with $\rho \sim \Sigma_{tot,0}/2h$ were to fragment hierarchically into successively denser structures, the entire conversion of gas into stars would take no longer than a time on the order of $\sim t_{ff,50}$, since the freefall time at all smaller scales is less than this. Counting the time for the initial growth of the gravitational instability, and the eventual gas evacuation due to feedback, we expect the entire period of star formation to last no longer than several freefall times \citep[e.g.][]{elmegreen:2000,elmegreen:2007}. This appears to be the case for Milky Way GMCs, which have a mean star-forming lifetime of 3 freefall times \citep{murray:2010.sfe.mw.gmc,lee:2016.gmc.eff}, as well as those found in simulated galaxies with low-temperature cooling and stellar feedback \citep{hopkins:fb.ism.prop}, however it has also been argued that star formation should take longer \citep{tan:2006,krumholz:sf.eff.in.clouds}.

\subsection{Star formation efficiency}

As stars form, the stellar surface density $\Sigma_\star(t)$ increases as the gas surface density $\Sigma_{gas}(t)=\Sigma_{tot,0} - \Sigma_\star(t)$ decreases. These stars will inject energy and momentum into the gas through various feedback mechanisms, however if the time-scale of star formation is so short that SNe do not occur then direct ISM heating can be neglected due to the short cooling time. Assuming that the stellar population is well-sampled from a \citet{kroupa:imf} IMF, the rate of momentum feedback injection per unit stellar mass $\frac{\dot{P}_\star}{m_\star}$ will initially be roughly constant, dominated by radiation pressure and fast winds from the most massive stars for the first $\unit[3]{Myr}$ after the stellar population forms. For the subsequent $\sim \unit[40]{Myr}$, the massive stars all leave the main sequence and supernovae become the dominant form of feedback. Because we are most interested in the limit of dense systems with short dynamical times, we can neglect stellar evolution and approximate $\frac{\dot{P}_\star}{m_\star}$ as being constant. Then the force of feedback upon the gas in the disk is:
\begin{equation}
F_{fb}(t) = \frac{\dot{P}_\star}{m_\star} M_\star =  \frac{\dot{P}_\star}{m_\star} \Sigma_\star(t) \pi R^2,
\label{FeedbackForce}
\end{equation}
assuming no leakage, photon trapping, or other effects arising from clumpy structure. This force will continue to increase until $F_{fb}$ exceeds the force of gravity binding the gas to the disk. The majority of the new star formation will occur in a thin disk, so while the gas is dense enough to form stars the gravitational field binding gas to the star-forming region will be dominated by contributions from the gas itself and the newly-formed stars. Thus:
\begin{equation}
F_g(t) =  g M_{gas}(t) = 2 \pi G \Sigma_{tot,0} \Sigma_{gas}(t) \mathrm{\pi} R^2.
\label{Gravity}
\end{equation}By equating the force of feedback upon the gas (\ref{FeedbackForce}) with that of gravity (\ref{Gravity}) we can determine the final stellar mass and hence the integrated star formation efficiency \citep{fall:2010.sf.eff.vs.surfacedensity}:
\begin{equation}
\epsilon_{int} = \frac{M_\star}{M} =  \frac{\Sigma_{tot,0}}{\Sigma_{tot,0} + \Sigma_{crit}}, 
\label{SFEformula}
\end{equation}
where:
\begin{equation}
\Sigma_{crit} = \frac{1}{2 \pi G} \frac{\dot{P}_\star}{m_\star}
\end{equation}
is the quantity with units of surface density encoding the strength of feedback relative to gravity. The contributions to $\frac{\dot{P}_\star}{m_\star}$ from radiation pressure, stellar winds, and SNe ejecta (ignoring the work done in the energy-conserving phase) are all of order $10^3 \frac{L_\odot}{\msun c}$. Thus, $\Sigma_{crit} \sim \unit[10^{3-4}]{\msun pc^{-2}}$ due to stellar feedback physics. Observationally, the average $\epsilon_{int}$ for Milky Way GMCs is $\sim 3\%$ \citep{murray:2010.sfe.mw.gmc,lee:2016.gmc.eff}, while the median GMC surface density is $\sim \unit[100]{\msun pc^{-2}}$ \citep{larson:gmc.scalings,solomon:gmc.scalings,bolatto:2008.gmc.properties}, so we can estimate that $\Sigma_{crit} = \unit[3000]{\msun pc^{-2}}$ for those GMCs for which feedback from massive stars is important. See also \citet{murray:molcloud.disrupt.by.rad.pressure}, \citet{dekel:2013.giant.clumps}, and \citet{thompson:2016.eddington.outflows} for similar derivations with various cloud and feedback models.

Equation (\ref{SFEformula}) predicts that the efficiency of starbursts occurring over adequately short time-scales is simply dictated by the ratio of forces of feedback and gravitation. In the limit $\Sigma_{tot,0} << \Sigma_{crit}$, SFE is proportional to $\epsilon_{int} \propto \Sigma_{tot,0}$ with the constant of proportionality determined by the strength of feedback. Inversely, where $\Sigma_{tot,0} >> \Sigma_{crit}$, SFE should approach unity: gravity prevails against feedback and converts nearly all gas to stars. The importance of surface density in determining star formation efficiency in short dynamical time systems is not simply a consequence of the `diskiness' of star-forming systems, nor of their optical depth in some band, both of which would give surface density an obvious physical relevance. It is merely a consequence of the fact that the ratio between the force of self gravity $F_{g} \sim \frac{G M^2}{R^2}$ and the momentum injection rate of feedback $F_{fb} \sim M_{\star} \nicefrac{\dot{P}_\star}{m_\star} $ has dimensions of surface density, at least under our simplifying assumptions. 

\section{Simulations}
\label{sec:sims}
Our simulations use {\small GIZMO} \citep{hopkins:gizmo}\footnote{A public version of this code is available at \gizmourl.}, a mesh-free, Lagrangian finite-volume Godunov code designed to capture advantages of both grid-based and smoothed-particle hydrodynamics (SPH) methods, built on the gravity solver and domain decomposition algorithms of {\small GADGET-3} \citep{springel:gadget}. In \citet{hopkins:gizmo} and \citet{hopkins:gizmo.mhd} we consider extensive surveys of test problems in both hydrodynamics and MHD, and demonstrate accuracy and convergence in good agreement with well-studied regular-mesh finite-volume Godunov methods and moving-mesh codes \citep[e.g.\ {\small ATHENA} \&\ {\small AREPO};][]{stone:2008.athena,springel:arepo}. We run {\small GIZMO} in its Meshless-Finite Mass (MFM) mode but have verified that Meshless Finite-Volume (MFV) mode produces nearly identical results (as expected from the previous studies). 

\subsection{Cooling, Star Formation, and Stellar Feedback}
\label{coolingSF}
The simulations here use the physical models for star formation and stellar feedback developed for the Feedback In Realistic Environments (FIRE) project \citep{hopkins:2013.fire,fire2}, although the simulations in this paper are idealized cloud collapse experiments on small scales, at often much higher mass resolution than the FIRE simulations. In general, we expect these methods to be appropriate to the scales examined in this work because by construction the FIRE framework adopts a physics approach that requires no phenomenological tuning to different mass scales. Hydrodynamics, gravity, cooling, and stellar feedback are explicitly and approximately solved down to the resolution limit, and the physics approximations invoked have been extensively validated by more expensive and detailed simulations. We briefly summarize some key properties of the FIRE models here, but refer to \citet{fire2} for details of the numerical implementations and extensive tests of the algorithms and physics.

When simulating gas fragmentation, it is critical to have explicit cooling physics; we therefore do {\em not} adopt an ``effective equation of state'' \citep{springel:multiphase} as has been done in many works in the past, but explicitly follow a wide range of heating/cooling processes. This includes photo-ionization and photo-electric, dust collisional, Compton, metal-line, molecular, and fine-structure processes, and we self-consistently account for optically thick cooling when local regions become thick to their own cooling radiation, implementing the approximation of \citet{rafikov:2007.convect.cooling.grav.instab.planets}. We do neglect the effects of non-equilibrium chemistry in the ISM, which can be very important for predictions of observational tracer abundances \citep{richlings:2014a,richlings:2014b}, however cooling times are generally so short in our problem that little dynamical effect can be expected.

Gas particles are converted to star particles with constant probability per unit time $t_{ff}\left(\rho\right)^{-1}$ if they satisfy all of the following star formation criteria:
\begin{itemize}
\item {\it Self-shielding and molecular:} We compute the molecular fraction $f_{mol}$ of the gas as a function of column density and metallicity according to \citet{krumholz:2011.molecular.prescription}, estimating the local gas column density with a Sobolev-like estimator.
\item {\it Contracting:} Star formation occurs only in regions of increasing density ($\nabla \cdot \vec{v}<0$).
\item {\it Self-gravitating:} The local Jeans mass $M_{jeans}$ is estimated, accounting for both turbulent \citep{hopkins:virial.sf} and thermal contributions, with the turbulent contribution typically dominating in cold molecular gas. Star formation is allowed only in regions where the Jeans mass can no longer be resolved, as it is at this point that fragmentation should continue down to unresolved scales.
\end{itemize}
In our tests, we find that the self-gravity criterion is the most restrictive and the most physically motivated of the above. Note that these criteria are slightly different from the FIRE simulations \citep{hopkins:2013.fire,fire2}, as we do not enforce a threshold density for star formation, and require gas to be increasing in density to form stars. All star formation criteria are fully adaptive, with no built-in scales that could be imprinted upon the star clusters that form. To summarize, gas fragmentation is explicitly followed down to the scale where the mass resolution is insufficient to resolve fragmentation, then the gas particles quickly (within one local $t_{ff}$) transition into collisionless star particles.

Crucially, because the collapse time-scale of {\em resolved} fragments at densities much larger than the mean in our simulations is always fast compared to the global dynamical time, this is not the rate-limiting step for star formation. Rather, it is the initial formation of these fragments \citep{thompson:rad.pressure, cafg:sf.fb.reg.kslaw, ostriker.shetty:2011}. As such, we will show that the star formation histories are insensitive to details of both our cooling and star formation prescriptions. This is consistent with a wide range of previous studies on GMC and galactic scales \citep{saitoh:2008.highres.disks.high.sf.thold, hopkins:rad.pressure.sf.fb, hopkins:fb.ism.prop, hopkins:stellar.fb.winds, hopkins:qso.stellar.fb.together,fire2,agertz:2013.new.stellar.fb.model}. 

Once stars form, feedback is included in the form of radiation pressure (UV, optical, and IR), stellar winds (fast, young star winds and slow AGB winds), SNe (types Ia and II), photo-ionization and photo-electric heating. Every star particle is treated as a single stellar population with an age based on its formation time and metallicity and mass inherited from its parent gas particle. Feedback includes the relevant mass, metal (with $11$ separately tracked species), momentum, and energy injection to the neighboring gas; all of the relevant quantities (stellar luminosities, spectral shapes, SNe rates, wind mechanical luminosities, yields) for the mechanisms above are tabulated as a function of time directly from the stellar population models in {\small STARBURST99}, assuming a \citet{kroupa:imf} IMF. For SNe, if we lack the mass resolution to resolve the Sedov-Taylor phase, we estimate the work done during the energy-conserving phase and couple the appropriate momentum based on fits from high-resolution SNR simulations (\citet{martizzi:2015.snr.inhomogeneous, kim:2015.snr.mom}, see \citet{hopkins:2013.fire} for implementation details). This is only important for our few simulations with resolved masses greater than $\unit[10^3]{\msun}$.

For the multi-band radiative fluxes necessary for the radiative heating and pressure terms, we use the LEBRON approximation, described in detail in \citet{fire2}. The spectrum is binned into UV, optical/near-IR, and mid/far-IR bands, and the approximate fluxes are computed explicitly at each particle. Local extinction around star particles is estimated with an effective column density computed with a Sobolev approximation; the robustness of our results to unknown order-unity factors in this prescription is demonstrated in Appendix \ref{appendix:radtest}. We emphasize that, unlike the model of \citet{hopkins:fb.ism.prop}, LEBRON does not invoke a subgrid ``boost'' term for the radiation pressure of multiply-scattered IR photons. Only explicitly-resolved photon absorption is accounted for in the heating and pressure terms.

We intentionally assign IMF-averaged properties to all star particles, rather than attempting to follow individual stars explicitly -- our goal is to study the effects of feedback, given some IMF, {\em not} to solve the problem of the origins and nature of the IMF itself. The latter would require a full model for individual star formation (and much higher resolution than we are able to achieve here), and may critically depend on additional physics (e.g.  heating by prostellar accretion, protostellar jets) which are negligible in an IMF-averaged feedback scenario. \footnote{One might worry that, by IMF-averaging, we make feedback ``too smooth.'' In limited experiments, we have crudely modeled the effects of stochastic sampling of the IMF and concentrating feedback in individual massive stars by, for each star particle, drawing from the IMF a quantized number of massive O-stars (from a Poisson distribution with mean equal to the expectation for the total mass of the particle). All feedback effects associated with massive stars (Type-II SNe, photo-heating, fast winds, radiation pressure) are multiplied appropriately by the number of O-stars (which are lost in each Type-II SNe event). As expected, this has essentially no effect on the disk-averaged properties we consider here for disk masses $\gtrsim 1000\,M_{\sun}$, which reasonably sample massive ($\gtrsim 10\,M_{\sun}$) stars. For still smaller clouds, this (as expected) introduces additional scatter in the star formation efficiency, corresponding to the variation in the number of massive stars (hence strength of feedback). However, the mean scalings are unaffected.} In some of our less-massive simulated clouds, the particle mass is less than $\msun$ and the stellar IMF is nominally resolvable, so star formation tends to produce ``clusters'' of star particles of $100\msun$ or less, which can be identified with the individual stars that would have formed. In this case, a sink-particle method (e.g. \citet{bate:1995.protobinary.accretion.vs.frag}) is certainly much more realistic and efficient, however we still adopt the standard star-particle method for consistency with the more massive clouds.

\begin{table*}
{\bf Simulation parameters}
\centering
\begin{tabular}{l|l|l|l|l|l|l}
\hline
 $\Sigma_{tot,0}$ $[\mathrm{M_\odot\,pc^{-2}}]$  & $R$ [$\mathrm{pc}$] & $M$ [$\mathrm{M_\odot}$] & $t_{ff,0}$  $[\unit[]{Myr}]$& Modifications & Mass Resolution $[\unit[]{\msun}]$  & Minimum star particle softening $[\unit[]{pc}]$ \\
(1) & (2) & (3) & (4) & (5) & (6) & (7) \\ \hline
127     & 5     & $10^4$ & 1.85        &       & 0.03   & 0.001         \\
127     & 50    & $10^6$ & 5.86       &       & 3      & 0.01  \\
127     & 500   & $10^8$& 18.53        &       & 300    & 0.1   \\ \hline
382     & 5     & $3\times10^4$ & 1.07        &       & 0.03   & 0.001         \\
382     & 50    & $3\times 10^6$ & 3.38       &       & 3      & 0.01  \\
382     & 500   & $3\times 10^8$ & 10.70        &       & 300    & 0.1   \\ \hline
1270    & 5     & $10^5$        & 0.59      &       & 0.1    & 0.001         \\
1270    & 50    & $10^7$ & 1.85       & ``Standard''      & 10     & 0.01  \\ \hline
1270    & 50    & $10^7$ & 1.85       & Random IC seeding 2   & 10     & 0.01  \\
1270    & 50    & $10^7$ & 1.85       & Random IC seeding 3   & 10     & 0.01  \\
1270    & 50    & $10^7$ & 1.85       & Optically-thin cooling        & 10     & 0.01  \\
1270    & 50    & $10^7$ & 1.85       & No feedback   & 10     & 0.01  \\
1270    & 50    & $10^7$ & 1.85       & $\nicefrac{1}{2}$-strength feedback   & 10     & 0.01  \\
1270    & 50    & $10^7$ & 1.85       & $\times 2$-strength feedback   & 10     & 0.01  \\
1270 	& 50	& $10^7$& 1.85		 & Radiation pressure only & 10 & 0.01 \\
1270    & 50    & $10^7$ & 1.85       & $150^3$ particle resolution   & 2.96     & 0.01  \\
1270    & 50    & $10^7$ & 1.85       & $50^3$ particle resolution    & 80     & 0.01  \\
1270    & 50    & $10^7$ & 1.85       & $1\%$ local SFR       & 10     & 0.01  \\
1270    & 50    & $10^7$ & 1.85       & $0.01Z_\odot$ initial metallicity     & 10     & 0.01  \\ \hline
1270    & 500   & $10^9$ & 5.86   &       & 1000   & 0.1   \\ \hline
3820    & 5     & $3\times 10^5$        & 0.34      &       & 0.3    & 0.001         \\
3820    & 50    & $3\times 10^7$ & 1.07        &       & 30     & 0.01  \\
3820    & 500   & $3 \times 10^9$ & 3.38       &       & 3000   & 0.1   \\ \hline
12700   & 5     & $10^6$ & 0.19     &       & 1      & 0.001         \\
12700   & 50    & $10^8$ & 0.59      &       & 100    & 0.01  \\
12700   & 500   & $10^{10}$ & 1.85       &       & 10000  & 0.1   \\ \hline
\end{tabular}
\caption{Initial conditions, numerical parameters and modifications of the simulations in this paper: (1): $\Sigma_{tot,0}$: the initial average gas surface density in $\unit[]{\msun\,pc^{-2}}$. (2): $R$: the radius of the initial spherical gas cloud in $\unit[]{pc}$. (3):  $M$: the initial gas mass in $\msun$.  (4): The freefall time  $t_{ff,0}$ at the initial density, defined in equation \ref{tff0}. (5): Modifications to the simulation with respect to the standard setup described in Section \ref{sec:sims}.  (6): Particle mass resolution in $\msun$. (7) Minimum Plummer-equivalent force softening for star particles. No minimum softening for gas particles is imposed. The particle number is $100^3$ in all simulations unless otherwise specified. All simulations start with solar metal abundances (except where stated otherwise), and an initial temperature of $\unit[10^4]{K}$.}
\label{simlist}
\end{table*}
\subsection{Initial Conditions \&\ Problem Setup}
The initial conditions of the simulations consist of a constant density gas sphere of radius $R$ and mass $M$, with the parameter space of $R$ and $M$ tabulated in table \ref{simlist}. These values are chosen to cover a range of values of $\Sigma_{tot,0}$, which, for reasons discussed in Section \ref{sec:derivation}, we expect to roughly parametrize the overall behaviour of the system even at disparate spatial scales, masses, and dynamical times. 

The initial velocity field is a superposition of solid-body rotation about the origin and a random turbulent component. The rotational frequency is set to the gas ball's Keplerian frequency $\Omega_K = \left(\nicefrac{G M}{R^3}\right)^{\frac{1}{2}}$, so that the effective radius, and hence average surface density of the disk remains roughly constant \footnote{Note that assuming rotational support is not a realistic choice for simulating GMCs, which are generally supported by a shearing velocity gradient and turbulence. As such, the simulations are not expected to result in large-scale cloud morphologies resembling realistic galactic GMCs. However, the morphology of sub-clouds will be determined on much shorter time-scales by local turbulence and self-gravity, independently of the large-scale morphology.}. The random velocity component adds a turbulent energy of $10\%$ of the initial gravitational binding energy, with a power spectrum $E(k) \propto k^{-2}$. All velocity Fourier coefficients for which $\|\vec{k}\|\geq\frac{2\pi}{R}$ are given a random phase and scaled according to this relation. The velocity components are first computed on a Cartesian grid circumscribing the gas sphere, and are then  interpolated to the particle positions.

The seed magnetic field is constructed in a similar fashion, such that the power spectrum of magnetic energy is also proportional to $k^{-2}$.The only difference from the above is that the $\nabla \cdot \vec{B}$ constraint is enforced by first computing random Fourier coefficients for the magnetic potential $\vec{A}$ and then applying the curl operator in Fourier space before transforming to real space in the same fashion as the velocity. The total magnetic energy is $1\%$ of the gravitational binding energy, which is $10\%$ of the initial turbulent energy. This figure was chosen based upon observations  suggesting that MHD turbulence in GMCs is super-Alfv\'{e}nic \citep{troland:2012.gmc.zeeman}, supported by high-resolution MHD simulations showing that the supersonic turbulent MHD dynamo tends to saturate the magnetic energy to $1-10\%$ of the turbulent energy \citep{federrath:supersonic.turb.dynamo}.

The gas is initialized to a temperature of $\unit[10^4]{K}$, however the simulations' results are insensitive to this choice because the cooling time in all cases considered is orders of magnitude shorter than the dynamical time-scale. At the beginning of the simulation, the gas immediately cools rapidly to several tens of $\unit[]{K}$, as is typical of the cold, neutral phase of the interstellar medium.

All simulations except those noted in table \ref{simlist} have $10^6$ particles, giving a fixed mass resolution of $10^{-6}M$. As discussed in Appendix \ref{convergence}, the star formation histories of the simulations are insensitive to our mass resolution at or above this level.

\section{Results}
\begin{table*}
{\bf Global simulation results}

\begin{tabular}{|l|l|l||l|l|l|l|l|l|l|l|l}
\hline
$\Sigma_{tot,0}$ $[\mathrm{\msun pc^{-2}}]$ & $R$ $[\mathrm{pc}]$ & Modifications & $\epsilon_{int}$ & $T_{SF}$ $[\mathrm{Myr}]$ & $T_{SF}/t_{ff,0}$&  $T_{2\sigma}$ $[\mathrm{Myr}]$ & $T_{2\sigma} / t_{ff,0} $  &  $\langle \epsilon_{ff,50} \rangle_{t}$     &  $\sigma_{\log \epsilon_{ff,50}}$ $[\unit[]{dex}]$ \\ 
(1) & (2) & (3) & (4) & (5) & (6) & (7) & (8) & (9) & (10)  \\ \hline
127	& 5	& 	& 0.04	 & 1.34	 & 0.72	 & 1.75	 & 0.94	 & 0.02 & 0.34	 \\ 
127	& 50	& 	& 0.04	 & 7.19	 & 1.23	 & 8.83	 & 1.51	 & 0.02	 & 0.56	 \\ 
127	& 500	& 	& 0.06	 & 25.50	 & 1.38	 & 35.20	 & 1.90	 & 0.01	 & 0.55	 \\ \hline
382	& 5	& 	& 0.11	 & 0.95	 & 0.89	 & 1.16	 & 1.09	 & 0.09	 & 0.70	 \\ 
382	& 50	& 	& 0.10	 & 4.23	 & 1.25	 & 5.04	 & 1.49	 & 0.07	 & 0.42	 \\ 
382	& 500	& 	& 0.11	 & 12.02	 & 1.12	 & 18.06	 & 1.69	 & 0.04	 & 0.61	 \\ \hline
1270	& 5	& 	& 0.31	 & 0.77	 & 1.31	 & 0.81	 & 1.38	 & 0.11	 & 0.77	 \\
1270	& 50	& ``Standard"	& 0.32	 & 2.22	 & 1.20	 & 2.45	 & 1.32	 & 0.12	 & 0.79	 \\ \hline
1270	& 50	& No Magnetic Field	& 0.34	 & 2.44	 & 1.31	 & 2.57	 & 1.39	 & 0.08	 & 0.74	 \\ 
1270	& 50	& Strong Magnetic Field	& 0.30	 & 2.33	 & 1.26	 & 2.59	 & 1.40	 & 0.11	 & 0.66	 \\ 
1270	& 50	& No feedback	& 0.86+	 & 3.25+	 & 1.75+	 & 3.59+	 & 1.94+	 & 0.52	 & 0.62	 \\
1270	& 50	& $1/2$ strength feedback	& 0.52	 & 2.53	 & 1.36	 & 2.77	 & 1.50	 & 0.18	 & 0.56	 \\ 
1270	& 50	& $\times 2$ strength feedback	& 0.19	 & 2.54	 & 1.37	 & 2.63	 & 1.42	 & 0.10	 & 0.57	 \\ 
1270	& 50	& Radiation pressure only	& 0.36	 & 2.49	 & 1.34	 & 2.59 & 1.4	 & 0.10	 & 0.85	 \\ 
1270	& 50	& Optically-thin cooling	& 0.32	 & 2.23	 & 1.20	 & 2.43	 & 1.31	 & 0.13	 & 0.54	 \\ 
1270	& 50	& Slow subgrid SFR	& 0.30	 & 1.79	 & 0.97	 & 1.85	 & 1.00	 & 0.11	 & 1.03	 \\ 
1270	& 50	& $Z=10^{-2}Z_\odot$	& 0.35	 & 2.05	 & 1.11	 & 2.13	 & 1.15	 & 0.14	 & 0.75	 \\ 
1270	& 50	& Random Seeding 2	& 0.30	 & 2.06	 & 1.11	 & 2.32	 & 1.25	 & 0.11	 & 0.56	 \\ 
1270	& 50	& Random Seeding 3	& 0.28	 & 2.03	 & 1.10	 & 2.23	 & 1.20	 & 0.10	 & 0.63	 \\ 
1270	& 50	& $150^3$ particle resolution	& 0.26+	 & 1.98+	 & 1.07+	 & 2.12+	 & 1.15+	 & 0.10	 & 0.60	 \\ 
1270	& 50	& $50^3$ particle resolution & 0.33	 & 2.78	 & 1.50	 & 3.10	 & 1.67	 & 0.10	 & 0.37	 \\ \hline
1270	& 500	& 	& 0.31	 & 7.50	 & 1.28	 & 7.91	 & 1.35	 & 0.14	 & 0.83	 \\ \hline
3820	& 5	& 	& 0.49	 & 0.55	 & 1.61	 & 0.61	 & 1.81	 & 0.13	 & 0.51	 \\ 
3820	& 50	& 	& 0.51	 & 1.58	 & 1.48	 & 1.73	 & 1.62	 & 0.14	 & 0.48	 \\ 
3820	& 500	& 	& 0.50	 & 5.06	 & 1.50	 & 5.35	 & 1.58	 & 0.16	 & 0.50	 \\ \hline
12700	& 5	& 	& 0.63	 & 0.33	 & 1.76	 & 0.36	 & 1.95	 & 0.20	 & 0.50	 \\ 
12700	& 50	& 	& 0.65	 & 1.02	 & 1.74	 & 1.17	 & 1.99	 & 0.20	 & 0.47	 \\ 
12700	& 500	& 	& 0.64	 & 3.14	 & 1.69	 & 3.37	 & 1.82	 & 0.20	 & 0.73	 \\
\end{tabular}
\caption{Important global quantities predicted by the simulations. Values denoted with a `+' indicate a lower bound. (1-3) As Table \ref{simlist}. (4)  $\epsilon_{int}$, the integrated star formation efficiency (equation \ref{def:sfe}). (5) $T_{SF}$, the characteristic width of the peak in the star formation history (equation \ref{def:tsf}), in $\mathrm{Myr}$. (6) $T_{SF}$ in units of the initial freefall time $t_{ff,0}$. (7) $T_{2\sigma}$, the interval of time containing $95\%$ of star formation in $\mathrm{Myr}$. (8) $T_{2\sigma} $in units of the initial freefall time $t_{ff,0}$. (9) $\langle \epsilon_{ff,50} \rangle_{t} $, the time-averaged per-freefall SFE defined in terms of the median gas density. (10) $\sigma_{\log \epsilon_{ff,50}}$, the dispersion in $\log \epsilon_{ff,50}$ in $\unit[]{dex}$.}
\label{resultlist}
\end{table*}
\label{sec:results}
\begin{figure*}
\includegraphics[width=\textwidth]{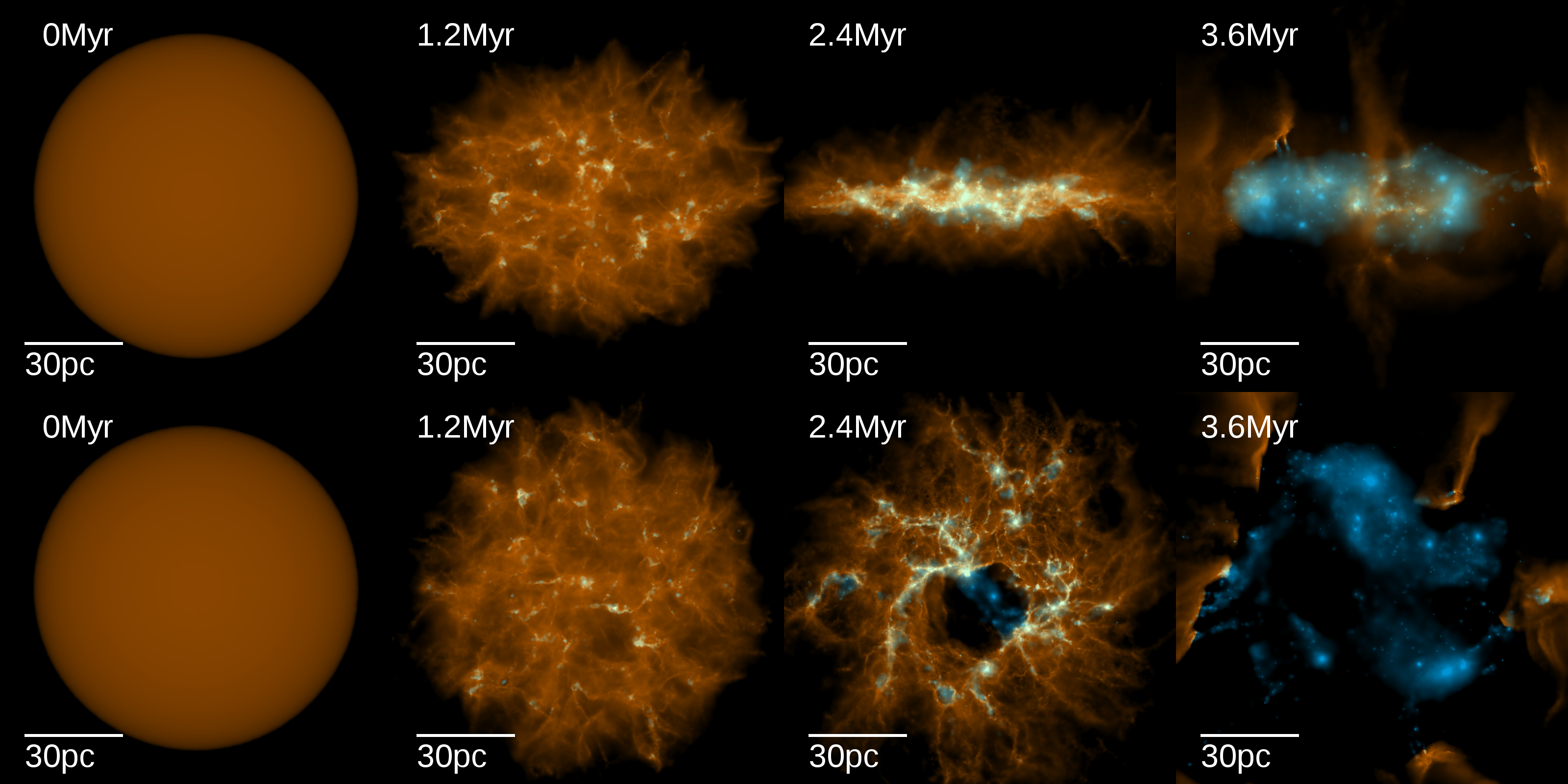}
\caption{Surface density of gas (orange) and stars (blue) in our fiducial run with parameters $M=3\times 10^7 \msun$ and $R = \unit[50]{pc}$, projected parallel ({\it top row}) and normal ({\it bottom row}) to the disk plane. {\it Far left:} The initial conditions, a uniform-density sphere. {\it Centre left:} After a time $\sim t_{ff,0}=\unit[1.2]{Myr}$, star formation has begun. {\it Centre right:} After another $t_{ff,0}$ has passed, the star formation rate has peaked and large star clusters have appeared. {\it Far right:} The system has reached the critical stellar mass, at which point the gas is blown out of the system by feedback, evacuating the central region.}
\label{prettypic}
\end{figure*}

\begin{figure*}
\centering
\includegraphics[width=\textwidth]{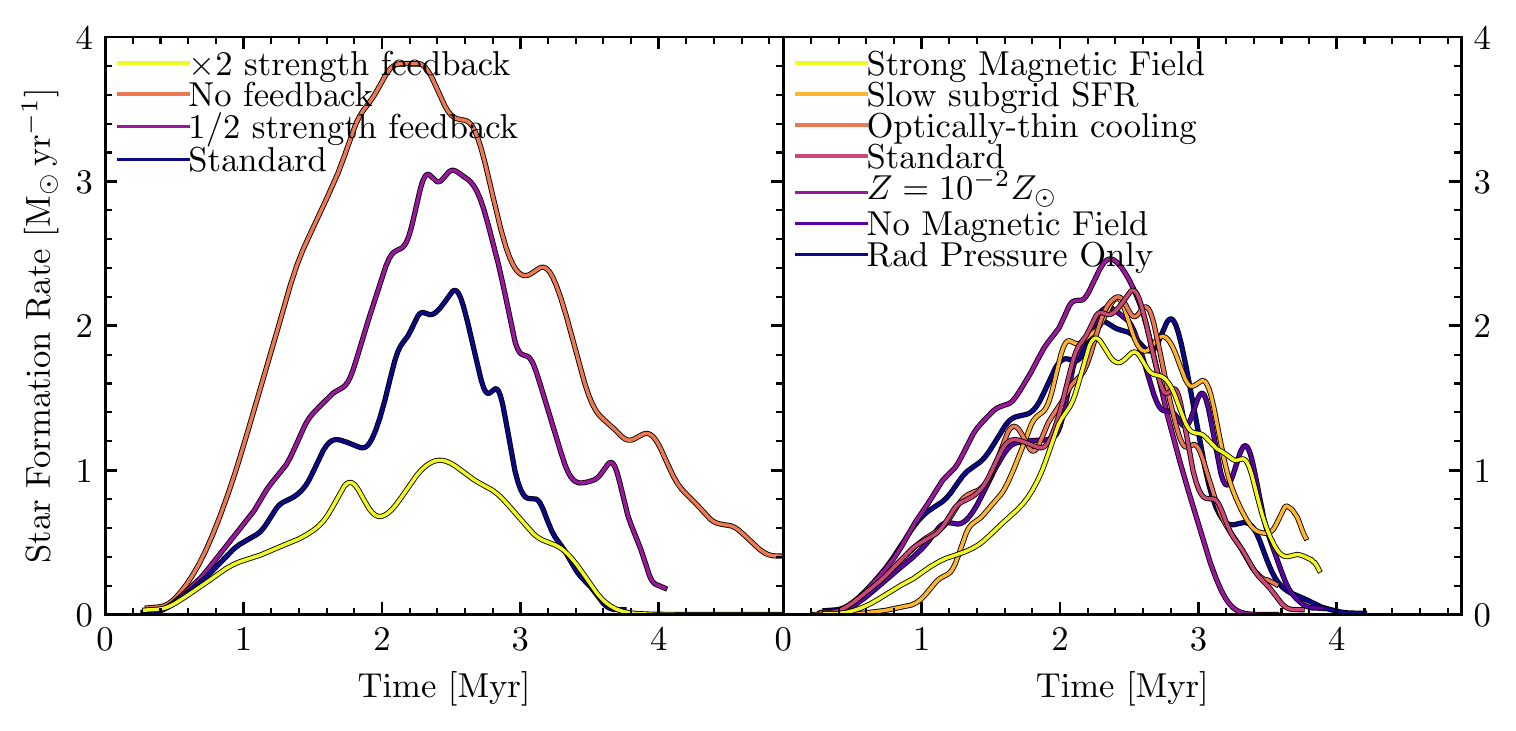}
\caption{Star formation histories of the physics test runs using the standard initial parameters $M=10^7\msun$ and $R=\unit[50]{pc}$. {\it Left:} Runs re-scaling the energy and momentum loadings of all stellar feedback mechanisms, producing large variations in the star formation history. {\it Right:} Our ``standard'' run compared to runs evolved from the same initial conditions with various physics options:  (1) {\it Strong magnetic field}: Setting the initial magnetic energy to $10\%$ of the binding energy, 10 times greater than standard. (2) {\it Slow subgrid SFR}: artificially ``slowing'' star formation in gas that satisfies the star formation criteria (Section \ref{coolingSF}) by multiplying the SFR by $\nicefrac{1}{100}$. (3) {\it Optically-thin cooling}: treating all radiative cooling as optically thin. (4) {\it $\mathit{Z=10^{-2}Z_\odot}$}: lowering the initial metallicity from $Z_\odot$ to $0.01Z_\odot$. (5) {\it No magnetic field}: turning off magnetic fields. (6) {\it Rad Pressure Only}: Removing all stellar feedback physics other than radiation pressure. These all produce relatively weak effects compared to simply rescaling the feedback energy and momentum fluxes, as discussed in section \ref{physicssurvey}}.

\label{CompareSFR}
\end{figure*}

\begin{figure*}
\includegraphics[width=\textwidth]{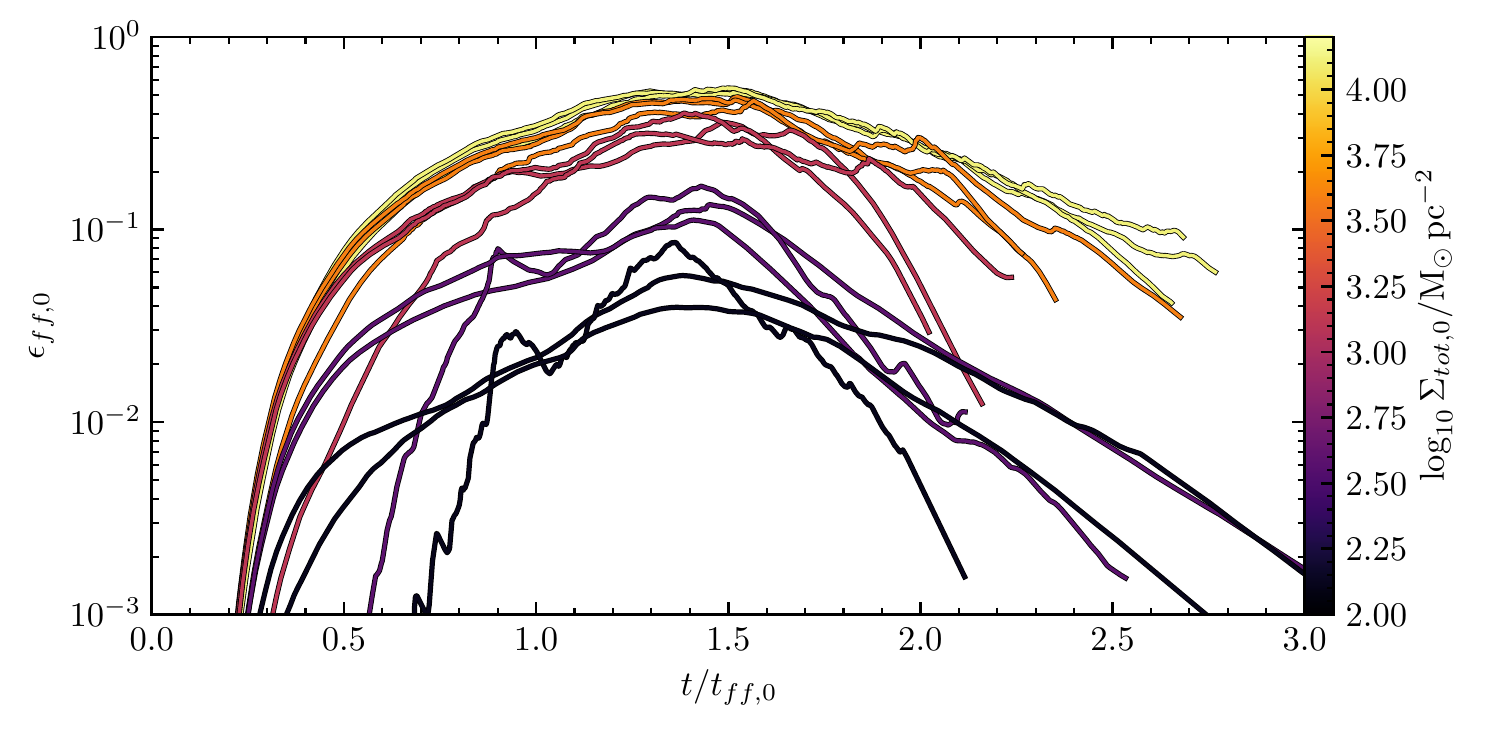}
\caption{Dimensionless star formation histories of all parameter survey runs:  the per-freefall SFE $\epsilon_{ff,0}=\frac{\dot{M}_\star t_{ff,0}}{M_{gas}}$ as a function of time in units of the initial freefall time $t_{ff,0}$ for the respective run. Each curve is a single run, coloured according to the value of $\Sigma_{tot,0}$. In all cases, $\epsilon_{ff,0}$ rises to a maximum dictated by the the strength of feedback relative to self-gravity, saturating to a value on the order of $1$ as $\Sigma_{tot,0}$ gets large.}
\label{eff_vs_t}
\end{figure*}

Qualitatively, all simulations follow the sequence of events illustrated in Figure \ref{prettypic}. The turbulent gas cloud immediately cools, with the lowest temperatures reaching $\sim \unit[10]{K}$. The initial velocity and magnetic fields seed density fluctuations and the gravitational instability grows, condensing the cloud into filaments and clumps. Within a freefall time, the first star clusters have formed. The star formation rate accelerates over $\sim \,t_{ff,0}$ to a peak value $\mathrm{SFR}_{max} \propto \epsilon_{ff} \nicefrac{M}{t_{ff}}$, with most star formation occurring in dense molecular sub-clouds. At this point the moderating effect of feedback comes into play and the SFR starts to drop as the disk acquires significant turbulent support. Eventually, all gas is blown out of the central region by feedback and star formation ceases. The product of the starburst is invariably a population of star clusters, some of which disperse upon gas expulsion, and some of which persist to the end of the simulation and remain bound. The end result is a population of star clusters surrounded by a diffuse, expanding gas shell.  

\subsection{Effects of Different Physics}
\label{physicssurvey}
In Figure \ref{CompareSFR}, we compare the star formation histories of the simulations evolved from identical initial conditions but with different physics enabled or disabled. It can be readily seen that the effect of varying the strength of feedback dwarfs all others, analogous to the conclusions of \citet{su:2016.feedback.first} for galaxy-scale star formation. Here we enumerate and describe these modifications and explain why, physically, this should be the case.
\subsubsection{Stellar feedback}
In one run, we neglect feedback altogether, and in two others we scale all energy and momentum feedback rates by $\nicefrac{1}{2}$ and $2$ respectively. We find that without any feedback moderation, star formation consumes nearly all ($86\%$ by the end of the simulation) gas within $\sim 2 t_{ff,0}$, with no sign of stopping. If the strength of feedback is scaled by $\nicefrac{1}{2}$, the star formation efficiency nearly doubles, while it is roughly halved when feedback is twice as strong, in agreement with equation \ref{SFEformula}. The time-scale for star formation remains unchanged, so the average per-freefall star formation efficiency $\epsilon_{ff}$ is also determined by the strength of feedback.

We also perform a run in which radiation pressure is the only feedback mechanism, and find that there is only maginally ($<10\%$) more star formation than the standard run. Thus, radiation pressure accounts for most of the feedback budget at this point in parameter space. We expect this to be generally true in clouds where the dynamical time does not greatly exceed $\unit[3]{Myr}$. Photoionization heating may have a significant contribution to disrupting the cloud if its escape velocity is $<\unit[10]{km\,s^{-1}}$ \citep{dale:2012}, but this will be the case for only a couple points in the parameter space of this paper. 

It is clear from the first panel of Figure \ref{CompareSFR} that the strength of feedback does not merely set the termination time of star formation: it also limits the star formation rate in an instantaneous sense - the stronger the feedback, the lesser the peak star formation rate. The specific feedback mechanism responsible for this is radiation pressure from young massive stars, as demonstrated by the radiation-pressure-only run. The radiation pressure is able to halt accretion onto cluster-forming cores, terminating star formation locally while it is still ongoing globally. Supernova feedback does not have this instantaneous effect due to its inherent time lag after initial star formation. Although we have not simulated it, a hypothetical starburst with only supernova feedback would proceed much like the zero-feedback run for the first $\unit[3]{Myr}$, which in this case is enough time to convert nearly all gas into stars. We therefore conclude that the early feedback mechanisms from massive stars are \emph{crucial} in setting the efficiency of rapid star formation in the high-density, short dynamical time regime studied in this work.

\subsubsection{Optically-thin cooling}
In one test run, we treating all radiative cooling as optically-thin (i.e. ignoring the optically-thick cooling suppression term from \citet{rafikov:2007.convect.cooling.grav.instab.planets}). This increases the cooling rate at high densities substantially. However, this has no discernible effect on the simulation results, as the opacity effects on the cooling function only become important in the suppression of fragmentation at the opacity-limited mass scale $\sim \unit[0.01]{\msun}$ \citep{rees:1976.opacity.limit}.

\subsubsection{Magnetic field strength}
We perform a simulation with no magnetic field and a simulation with a ``strong'' magnetic field whose initial magnetic energy is equal to the initial turbulent energy, 10 times the standard value. A strong enough magnetic field may suppress fragmentation and the local SFR by as much as a factor of 2 on small scales \citep{federrath:2012.sfr.vs.model.turb.boxes}, without considering feedback. We do see this effect in the ``strong'' magnetic field run: the initial star formation rate is about $\nicefrac{1}{2}$ that of the standard run. However, the SFR still continues to rise until it reaches the level set by feedback moderation, and the rest of the star formation history is quite similar to the other runs. Removing the magnetic field had no discernible effect upon the SFR, suggesting that the magnetic field has no large-scale dynamical relevance in the standard physics runs. However, we do note a small-scale cloud morphology in the MHD simulations that is distinctly more filamentary than the non-MHD simulation, due to the gas preferentially moving along magnetic field lines \citep[see][]{collins:2012.mhd.sf}.

\subsubsection{Slow subgrid SFR}
In this run, we force a small-scale star formation rate $\dot{\rho}_\star=0.01\rho_{mol}/ t_{ff}$ in gas that satisfies the star formation criteria (Section \ref{coolingSF}). This is 100 times slower than the usual choice, and comparable to the specific star formation rate on the scale of galactic disks \citep{kennicutt98,krumholz:2012.universal.sf.efficiency}. This does not affect the average SFR in our simulations because the rate-limiting step of star formation is the formation of dense, unstable gas structures in the first place. Collections of gas particles that meet the star formation criteria but have not yet turned into stars will simply continue to contract to greater densities within a local freefall time, causing the local SFR to diverge until stars inevitably form. This result is notably different from simulations which enforce the same star formation law but do {\it not} follow low-temperature cooling below $10^4 K$ and adopt an effective equation of state for stellar feedback. In such a simulation, the local star formation law would underestimate the global star formation rate because the aforementioned gravitational contraction would be suppressed.

Note that this insensitivity to the local star formation efficiency is only obtained because the gas particle gravitational softening is fully adaptive. Otherwise, the cold gas would simply contract to inter-particle spacings comparable to the minimum softening and stop at that density, and the local SFR would stop increasing.

The most notable effect of this modification was the formation of much denser and much more plentiful bound star clusters. As gas exhaustion is slowed down locally, protoclusters spend more time radiating away energy, contracting, and damping out their internal turbulent motions before turning into star particles. This increases the compactness and boundedness of the remnants. We therefore caution that while global star formation histories are not sensitive to the local value of $\epsilon_{ff}$ (see also \citealt{fire2}), the physics of star cluster formation may be.

\subsubsection{Metallicity}
In the low-metallicity test, we scale the initial gas metallicity down from $Z_\odot$ to $10^{-2}Z_\odot$. This can affect many aspects of the cooling and feedback physics. Metal line cooling is proportionally less efficient, however even at $Z \sim 10^{-2} Z_\odot$, $t_{cool}<<t_{ff}$ in the most dense gas, so fragmentation should not be strongly altered. This may change at metallicities of $10^{-4}-10^{-5}Z_\odot$ \citep{hopkins:2015.metal.poor}. The metallicity also determines dust opacity, and thus the coupling efficiency for IR radiation pressure. Lastly, it affects the evolution of the formed stellar populations' mass, energy and momentum injection rates, which are obtained from {\small STARBURST99}. Overall, the metal-poor simulation had a star formation efficiency only marginally greater than the standard run ($0.35$ compared to $0.32$), however it did have a faster initial growth in the SFR, suggesting that the stellar feedback at low metallicity might be less effective at halting accretion onto cluster-forming cores. The main difference in the feedback budget is due to the $\propto Z^{0.7}$ scaling of the line-driven stellar wind mass loss rate of type O stars \citep{vink:2001.ob.mass.loss}. At solar metallicity, the momentum input is somewhat less than that of radiation pressure, but the same order of magnitude. At $10^{-2} Z_\odot$, however, the dynamical effect of the winds is negligible.

We have also performed limited experiments with our routines for cosmic ray heating, cooling, streaming and diffusion. In general, if the system is given an initial cosmic ray energy density, it will rapidly cool away into dynamical irrelevance: like the magnetic field, it is ultimately a reservoir for the energies of gravitational collapse and stellar feedback, and not a source of energy in itself. There is also the possibility of the system being immersed in a strong cosmic ray background, however such environmental interactions are beyond the scope of this work. However, \cite{starburst.cosmic.rays} have found that the cosmic ray energy in nuclear starbursts tends to be considerably smaller than the magnetic field energy, suggesting that even in the full picture with a realistic galactic environment cosmic rays should not greatly influence the overall dynamics of a collapsing GMC.

\begin{figure*}
\centering

\includegraphics[width=\textwidth]{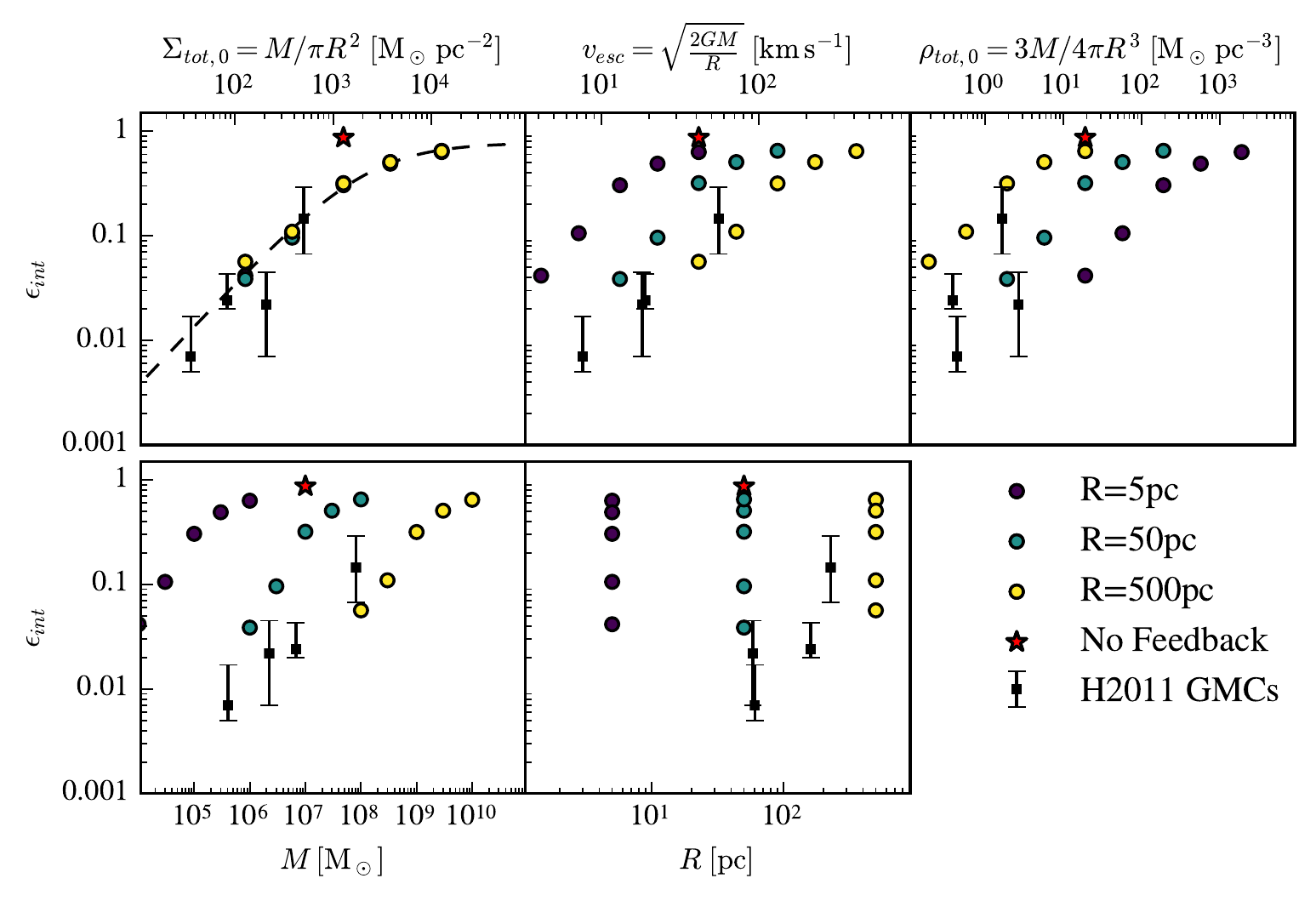}
\caption{Integrated SFE $\epsilon_{int}$ of the 15 parameter survey simulations plotted against various functions of the initial simulation parameters $M$ (mass) and $R$ (radius). The points with error bars, ``H2011 GMCs'', represent the populations of giant molecular clouds extracted from previous full-scale galaxy simulations \citep{hopkins:rad.pressure.sf.fb}. The points represent the population medians, and the bars represent the $\pm 1\sigma$ percentiles. The dashed line in panel 1 is the best-fit curves to equation \ref{SFEfit}, which gives parameters $\Sigma_{crit}=\unit[2800\pm100]{\msun\,pc^{-2}}$ and $\epsilon_{max}=0.77 \pm 0.05$.}
\label{SFEvSigma}
\end{figure*}

\subsection{Integrated star formation efficiency}
\label{sec:results:sfe}

We now arrive at our main results. In Figure \ref{SFEvSigma} the star formation efficiencies of the parameter survey simulations are plotted against the surface density, escape velocity, 3D density, mass and radius derived from the simulation parameters $M$ and $R$. Clearly, the mass, size, density, and escape velocity are {\it not} good general predictors of $\epsilon_{int}$; similar $\epsilon_{int}$ values are obtained in simulations for which these quantities differ by orders of magnitude. 

Of the obvious physical quantities derived from $M$ and $R$, $\Sigma_{tot,0}$ is the best predictor of $\epsilon_{int}$, with particularly good agreement between spatial scales at high $\Sigma_{tot,0}$, where the dynamical time is always short compared to main sequence lifetimes. In general, we obtain good agreement with equation \ref{SFEformula}: $\epsilon_{int}$ scales $\propto \Sigma_{tot,0}$ when $\Sigma_{tot,0}<< \Sigma_{crit}$, and it saturates to a maximum $\epsilon_{int}$ at sufficiently high surface density. The saturation efficiency is not necessarily 1, as depends on the initial conditions and what subset of the gas is used when defining $\epsilon_{int}$. As an extreme example, if the initial gas density field had an extended warm diffuse background component, as it might realistically, the diffuse gas would never form stars over the time-scale of interest, but would reduce the $\epsilon_{int}$ statistic if it were included in the gas mass sum. In our simulations, it is possible that there is a similar effect for the diffuse gas at the outer edges of the disk, as well as the gas which escapes through under-dense `chimneys' between the dense sub-clouds within the disk.

We fit $\epsilon_{int}$ to the following two-parameter model:
\be
\epsilon_{int} = \left( \frac{1}{\epsilon_{max}} + \frac{\Sigma_{crit}}{\Sigma_{tot,0}}\right)^{-1},
\label{SFEfit}
\ee

which is equivalent to the \citet{fall:2010.sf.eff.vs.surfacedensity} formula (equation \ref{SFEformula}) in the limit $\Sigma_{tot,0}<<\Sigma_{crit}$ but approaches $\epsilon_{max}$ as $\Sigma_{tot} \rightarrow \infty$. Performing an unweighted fit on $\log \epsilon_{int}$, the best-fit parameters are $\Sigma_{crit} = \unit[2800\pm 100]{\msun\,pc^{-2}}$ and $\epsilon_{max} = 0.77\pm 0.05$. The best-fit curve is plotted in panel 1 of Figure \ref{SFEvSigma}. This value of $\Sigma_{crit}$ is within a factor of 2 of that found by \citet{fall:2010.sf.eff.vs.surfacedensity}, and is compatible with the value of $\Sigma_{crit}$ found in Section \ref{sec:derivation} from the average observed $\epsilon_{int}$ of Milky Way GMCs.

The residual $R$-dependence of $\epsilon_{int}$ is small, but is positively correlated with $R$. This may be explained by the built-in scales in ISM cooling and stellar feedback physics. It is expected that the thermal pressure of the warm ISM heated to $\unit[10^4]{K}$ will have a greater proportional dynamical effect in the few clouds with escape velocities that do not greatly exceed $\unit[10]{km\,s^{-1}}$. The time-scale of stellar evolution also introduces a scale into stellar feedback: at fixed $\Sigma_{tot,0}$, $t_{ff}$ scales $\propto R^\frac{1}{2}$. Therefore, as $R$ spans $\unit[2]{dex}$, the time-scale of star formation spans an order of magnitude, so the timing of star formation relative to the stellar evolution within the formed stellar populations varies with $R$ at fixed $\Sigma_{tot,0}$. Stellar evolution causes $\frac{\dot{P}_\star}{m_\star}$ to vary over time, so the effective strength of feedback that determines $\epsilon_{int}$ will be some function of the global star formation time-scale $t_{ff}$. The general trend is that of increasing SFE over longer dynamical times, indicating that the effective $\frac{\dot{P}_\star}{m_\star}$ decreases monotonically with time. This is {\it despite} the increasing relevance of supernovae in the simulations spanning longer time-scales: as massive stars die, the introduction of supernovae is not enough to make up for the loss of mechanical luminosity from radiation and stellar winds to maintain the initial $\frac{\dot{P}_\star}{m_\star}$.

In Figure \ref{SFEvSigma}, the compiled SFE statistics for GMC populations extracted from the parameter survey of full-scale galaxy simulations \citep{hopkins:fb.ism.prop} are also plotted for comparison, and happen to be largely compatible with the fit. In light of this and the agreement with the observational estimate of $\Sigma_{crit}$, we may safely generalize these results from our contrived generic gas ball setup to clouds with actual GMC morphologies as they emerge from galactic gas dynamics. While the large-scale morphology and relative importances of shear, rotation, and turbulence may be different between our simulations and GMCs that emerge in galaxy simulations, the scaling of $\epsilon_{int}$ is an inevitable result that applies to self-gravitating gas cloud that can form stars. Therefore, equation \ref{SFEfit} is a general predictor of the $\epsilon_{int}$ of a star-forming gas cloud, provided that it is self-gravitating and it has some well-defined average surface density.

\subsection{Duration of star formation and per-freefall SFE}\label{section:results:eff}
\begin{figure*}
\includegraphics{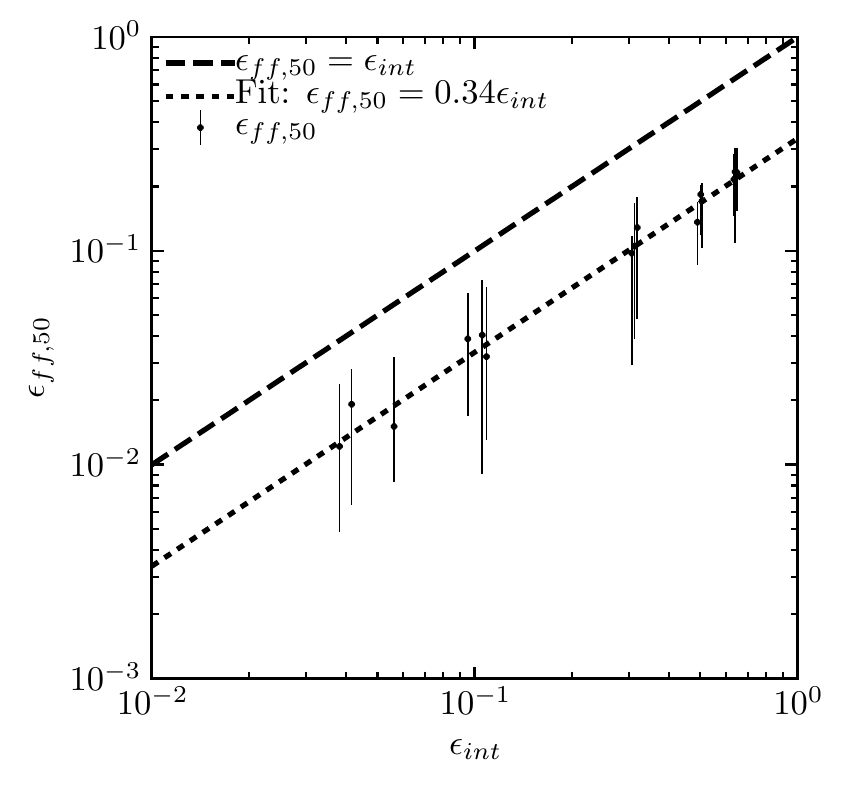}
\includegraphics{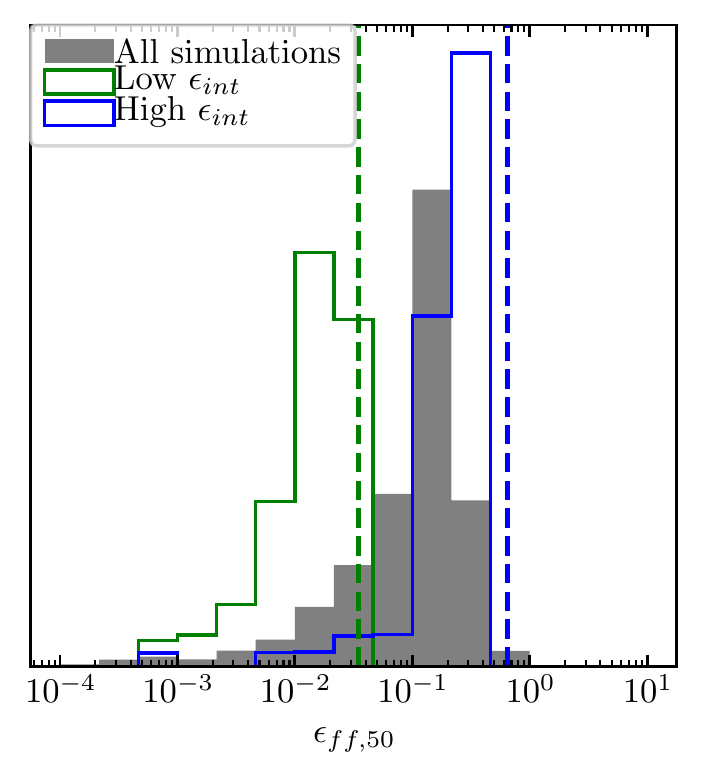}
\caption{{\it Left:} Instantaneous per-freefall star formation efficiency $\epsilon_{ff,50} = \dot{M}_\star \left( t \right)\, t_{ff,50} \left( t \right) / M_{gas}\left(t\right)$ (see equation \ref{def:tffmode}) as a function of integrated star formation efficiency $\epsilon_{int}$ for all parameter survey simulations. The points represent the value of $\epsilon_{ff,50}$ averaged over all times where the SFR is nonzero. Error bars represent the $\pm 1\sigma$ percentiles of $\epsilon_{ff,50}$. The dashed line marks the line of equality between $\epsilon_{ff}$ and $\epsilon_{int}$, and the dotted line indicates the best proportional fit. {\it Right:} Histograms of $\epsilon_{ff,50}$ for all parameter-survey simulations (grey), a highly-efficient ($\epsilon_{int}=0.64$) run with $\Sigma_{tot,0}=\unit[12700]{\msun\,pc^{-2}}$, $R=\unit[50]{pc}$ (blue), and an inefficient ($\epsilon_{int}=0.08$) run with $\Sigma_{tot,0}=\unit[382]{\msun\,pc^{-2}}$, $R=\unit[50]{pc}$ (green). The dashed lines indicate $\epsilon_{int}$ for the respective runs. Not surprisingly, $\epsilon_{ff,50}$ scales in proportion to $\epsilon_{int}$, but it has considerable variation ($\sim \unit[0.4-0.8]{dex}$) throughout the star formation history of a single simulation. For Milky Way GMCs of surface density $\sim \unit[100]{\msun\,pc^{-2}}$, we expect $\epsilon_{ff,50}$ to average to $0.01$, in good agreement with observations.}
\label{eff_vs_sfe}
\end{figure*}

We now discuss results concerning star formation rates and timescales.  As stated in the overview, star formation in all parameter survey simulations spans no more than $\sim 3t_{ff,0}$ (see Figures \ref{CompareSFR} and \ref{eff_vs_t}). Here we seek to quantify this statement more precisely. As a general-purpose measure of the duration of the starburst, we define the quantity $T_{SF}$, the stellar mass formed divided by the mass-weighted average star formation rate:
\begin{equation}
T_{SF} = \frac{M_\star}{\langle \dot{M_\star}\rangle} = \frac{M_\star^2}{\int \left(\dot{M_\star}\right)^2\,\mathrm{d}t}.
\label{def:tsf}
\end{equation}
This is a natural measure of characteristic of the peak in the star formation history (see Figures \ref{CompareSFR} and \ref{eff_vs_t}). It is also a useful proxy for the lifetime of the gas disk, as star formation largely begins once the gas has settled into a disk and halts once the disk is disrupted. The values of $T_{SF}$ are tabulated in table \ref{resultlist}. $T_{SF}$ is insensitive to the small early and late tails of the star formation history, however, so in table \ref{resultlist} we also quote $T_{2\sigma}$, the time interval containing $95\%$ of the star formation. This is generally only slightly more than $T_{SF}$, as most star formation occurs in a brief burst, and feedback is able to rapidly quench star formation.

In all simulations, $T_{SF} \sim t_{ff,0}$ (see table \ref{resultlist}), so most of the star formation occurs within a single initial global freefall time. This confirms our argument in Section \ref{sec:derivation}: since $t_{ff,0}$ is longer than any other internal collapse time-scale, and turbulent support dissipates in a crossing time \citep[e.g.][]{elmegreen:2000}, the disk should be able to form enough stars to reach the blowout stage within this time. This time constraint implies a tight relation between $\epsilon_{int}$ and $\epsilon_{ff}$: if star formation is constrained to happen over $N$ dynamical times, then $\epsilon_{ff} = \epsilon_{int} /N$ on average.

This brings us to a very important subtlety of feedback-moderated star formation: while stellar feedback determines $\epsilon_{int}$ in a simple way through the force balance described in Section \ref{sec:derivation}, it also determines $\epsilon_{ff}$ in an ``instantaneous'' sense, with ``instantaneous'' meaning over time-scales much longer than the dynamical time of the smallest resolved units of star formation, yet still much shorter than the global timescale. Since star formation is a process of hierarchical fragmentation from the largest cloud scale down to individual stars, the total star formation history is the sum of a hierarchy of many individual smaller and shorter star formation events, each of which has its $\epsilon_{int}$ determined by the local ratio of feedback and gravity. This results in an overall star formation rate that is moderated ``from the bottom up". Realistically, the ``bottom'' of this hierarchy would be set by the mass scale at which it is likely that the sampled IMF contains a massive star that can exert strong feedback.

It is of limited usefulness to compare star formation time-scales to $t_{ff,0}$, at least when comparing with the value of $\epsilon_{ff}$ in observed star-forming systems, as it requires knowledge of the more-diffuse initial conditions. The freefall time inferred for the gas disks as they would be observed during star formation would be something closer to $t_{ff,50}$, as derived from the mass-weighted median gas density (Equation \ref{def:tffmode}) \footnote{We have found that in these simulations $t_{ff,50}$ tends to be quite close to the freefall timescale derived from the volume-averaged gas density, which is closer to what is actually calculated for GMCs. We use $t_{ff,50}$ because we have found it to be more stable and robust.}. Average values of $\epsilon_{ff,50} \equiv \dot{M}_\star \left( t \right)\, t_{ff,50} \left( t \right) / M_{gas}\left(t\right)$  for each simulation can be found in columns 9 and 10 of table \ref{resultlist}. In panel 1 of Figure \ref{eff_vs_sfe} we plot $\epsilon_{ff,50}$ as a function of $\epsilon_{int}$ and confirm that there is a tight relation between two efficiencies. The best-fit power law to the relation has an exponent within $1\sigma$ of 1, so we propose a simple proportional relation:
\begin{equation}
\langle \epsilon_{ff,50} \rangle_t = 0.34 \epsilon_{int},
\label{eff_vs_eint}
\end{equation}
where $\langle \epsilon_{ff,50} \rangle_t$ denotes the average observed value at a random point during the star formation history. The physical implication of this relation is that star formation in the simulations is indeed constrained to occur mainly within $\sim 3$ dynamical times, regardless of the relative strength of feedback and gravity, as was argued in Section \ref{sec:derivation}. This would agree with the mean GMC lifetime of 3 freefall times inferred in \citet{murray:2010.sfe.mw.gmc}.

The shape of the distribution of $\epsilon_{ff}$, which we show in panel 2 of Figure \ref{eff_vs_sfe}, is also of interest. In general, the distribution is strongly peaked near $\epsilon_{int}$, with only brief excursions above $\epsilon_{int}$. The distribution is negatively skewed due to the early and late tails of the star formation history, which spread the distribution over several orders of magnitude, similar to what is found in \citep{lee:2016.gmc.eff}. The intrinsic dispersion in the value of $\epsilon_{ff,50}$ across the lifetime of the system (Table \ref{resultlist}, Column 10) typically has a value between 0.4 and $\unit[0.8]{dex}$.

\section{Discussion}
\label{sec:discussion}
\subsection{Star-forming clouds and clumps in the Milky Way}
\label{observations}
Many star-forming clouds, identified as associations between emission from young stars and molecular gas, have been observed in the Milky Way. These clouds can be broadly classified into two groups: GMCs proper, which have characteristic surface density $\unit[100]{\msun\,pc^{-2}}$ and are typically traced in CO \citep{larson:gmc.scalings,solomon:gmc.scalings, bolatto:2008.gmc.properties}, and dense clumps, which have a typical surface density of $\unit[10^3]{\msun\,pc^{-2}}$, and are traced in higher-density tracers such as HCN \citep{wu:2005.clumps,wu:2010.clumps,heyer:2016.clumps}. The observational proxy of $\epsilon_{int}$ that can be obtained for these systems is:
\begin{equation}
\epsilon_{obs} = \frac{M_{\star,young}}{M_{\star,young}+M_{molecular}},
\label{eq:epsobs}
\end{equation}
where $M_{\star,young}$ is the mass of stars younger than $\unit[3.9]{Myr}$, as can be traced from emission from HII regions or from direct counts of young stellar objects, and $M_{molecular}$ is the mass of molecular gas in the cloud. Note that both of these masses must vary during a star-forming cloud's lifetime, and in general $\epsilon_{obs} \neq \epsilon_{int}$. However, the trend in $\epsilon_{obs}$ with $\Sigma_{gas}$ should still follow that of $\epsilon_{int}$, so some systematic variation in the $\epsilon_{obs}$ should be evident in clouds with widely different surface densities.

In Table \ref{table:obs}, we summarize the $\Sigma_{gas}$ and $\epsilon_{obs}$ statistics of the GMC datasets of \citet{lee:2016.gmc.eff} and \citet{vuti:2016.gmcs} and the dense clump datasets of \citet{wu:2010.clumps} and \citet{heyer:2016.clumps}. \citeauthor{lee:2016.gmc.eff} provides $\epsilon_{obs}$ directly (denoted $\epsilon_{br}$ in the paper). We estimate $M_{\star,young}$ from the \citeauthor{vuti:2016.gmcs} dataset by multiplying the provided SFR measurements from MIR flux by $\unit[3.9]{Myr}$, the mean massive star lifetime weighted by ionizing flux \citep[e.g.][]{murray:2010.sfe.mw.gmc} . We compute $M_{\star,young}$ in the \citeauthor{wu:2010.clumps} clumps by converting the reported IR luminosities to the mass of a young single stellar population with a \citet{kroupa:imf} IMF. In Table \ref{table:obs} we give values for the \citeauthor{heyer:2016.clumps} corresponding to the value of $M_{\star,young}$ extrapolated from YSO counts assuming a \citeauthor{kroupa:imf} IMF (an upper bound) as well as values assuming the only mass is in stars that have been directly counted (a lower bound). As it is physically unlikely that less massive stars are not present, and the SFE from the upper bound is closer to \citet{wu:2010.clumps} and nearby star-forming regions \citep{lada:2003.embedded.cluster.review}, the true value is probably closer to the upper bound.

A $\unit[\sim1]{dex}$ scaling in the median $\epsilon_{obs}$ is evident between $\sim 1\%$ for the GMCs at $\sim\unit[10^2]{\msun\,pc^{-2}}$ and $\sim 10\%$ for the clumps at $\sim\unit[10^3]{\msun\,pc^{-2}}$, in agreement with the general prediction of our SFE model. However, substituting the median surface density into our model for $\epsilon_{int}$ (Equation \ref{SFEfit}) gives a SFE that is typically $\sim\unit[0.4]{dex}$ greater than the median $\epsilon_{obs}$. This offset could have several possible causes, including an underestimation of the strength of feedback in the simulations, the accounting of gravitationally bound gas in the observations, or an intrinsic bias in $\epsilon_{obs}$ as an estimator of $\epsilon_{int}$.

If the scatter in the observed $\epsilon_{obs}$ were only due to intrinsic variation from the scatter in $\Sigma_{gas}$, then we would expect the scatter in $\Sigma_{gas}$ and $\epsilon_{obs}$ to be equal. This is not the case: the scatter in $\epsilon_{obs}$ is too large to be explained by the variation in $\Sigma_{gas}$ alone. This is likely due to the variation in the observed $\epsilon_{obs}$ that arises from observing the clouds at random times in their star-forming lifetimes as the stellar and molecular mass content varies \citep[e.g.][]{lee:2016.gmc.eff}. This type of variation is present to some extent in the simulations (e.g. Figure \ref{eff_vs_sfe}, panel 2). 

\begin{table*}
\begin{tabular}{l|l|l|l|l}
Dataset & Class & $\log\Sigma_{gas}\,(\unit[]{\msun\,pc^{-2}})$ &  $\log\epsilon_{obs}$ & $\log \epsilon_{int}$ predicted from median $\Sigma_{gas}$ \\ \hline
\vspace{2pt} \citealt{lee:2016.gmc.eff} & GMCs & $1.88^{2.19}_{1.40}$ & $-1.97^{-1.23}_{-2.76}$ & -1.58\\ 
\vspace{2pt} \citealt{vuti:2016.gmcs} & GMCs & $1.95^{2.24}_{1.68}$ & $-1.93^{-1.37}_{-2.58}$ & -1.51\\
\vspace{2pt} \citealt{wu:2010.clumps} & Dense clumps & $3.00^{3.39}_{2.63}$ & $-1.10^{-0.86}_{-1.76}$& -0.61 \\
\vspace{2pt} \citealt{heyer:2016.clumps} & Dense clumps & $2.79^{3.05}_{2.61}$ & Upper: $-0.87^{-0.55}_{-1.29}$, Lower: $-2.14_{-2.71}^{-1.69}$ & -0.76 \\
\end{tabular}
\caption{Quantiles of molecular gas surface density $\Sigma_{gas}$ and the observationally-inferred SFE $\epsilon_{obs}$ (Equation \ref{eq:epsobs}) from various studies of star-forming GMCs and dense clumps in the Milky Way, in the format $\mathrm{median}^{+1\sigma}_{-1\sigma}$. Both $\Sigma_{gas}$ and $\epsilon_{obs}$ typically scale by $\sim\unit[1]{dex}$ between GMC conditions and dense clump conditions. For \citet{heyer:2016.clumps}, both upper and lower bounds are provided. The final column gives the true integrated SFE $\epsilon_{int}$ predicted by substituting the median $\Sigma_{gas}$ into Equation \ref{SFEfit}.}
\label{table:obs}
\end{table*}

We may also compare to observational estimates of $\epsilon_{ff}$. The \citeauthor{lee:2016.gmc.eff} and \citeauthor{vuti:2016.gmcs} datasets give median $\epsilon_{ff}$ values of $\sim 2\%$ and $\sim 1\%$ respectively, which are consistent with what is found in our simulations with similar gas surface density. However, the best-fit $\epsilon_{ff}$ in dense clumps reported by \citeauthor{heyer:2016.clumps} is also $\sim 2\%$ when the upper bound on the stellar mass is used. In \citet{heyer:2016.clumps}, the SFRs are computed by dividing the inferred stellar mass by $\tau_{SF}=\unit[0.5]{Myr}$, the evolution timescale for Class I protostars inferred from low-mass star-forming regions \citep{evans:2009.sfe,gutermuth:2009.ysos}. In general, inferred SFRs of dense clumps have relied on assumption that star formation has been steady for at least as long as $\tau_{SF}$\footnote{The SFRs of the \citet{wu:2010.clumps} HCN clumps, as determined by \citet{heiderman:2010.gmcs} from infrared luminosity, have $\epsilon_{ff}\sim 1\%$, but again there is an implicit averaging window $\tau_{SF} \sim \unit[4]{Myr}$ in the $L_{IR}-SFR$ conversion factor used. This figure of $1\%$ does appear to be a general finding for dense clumps \citep{krumholz:2014.review}.}, which is questionable within the picture presented in this paper given that nearly all clumps have freefall times shorter than this. If the lifetime of HCN clumps is significantly longer than $\unit[0.5]{Myr}$ (and they are as dense as presumed) it must be due to some physics that are is not accounted for in this work. One possibility is a transition in the nature of star-forming flows at lower Mach numbers, which we have hardly surveyed in our simulations. The clumps in \citeauthor{heyer:2016.clumps} have a characteristic velocity dispersion of $\sim \unit[0.75]{km\,s^{-1}}$, corresponding to a Mach number of $2-3$, much less supersonic than GMCs at large, and in the range expected from monolithic isothermal collapse \citep{larson:1969.isothermal.collapse,penston:1969.isothermal.collapse}. Such a transition in the nature of the flow below $\unit[1]{km\,s^{-1}}$ is suggested by the inverse size-linewidth relation of clumps \citep{wu:2010.clumps} compared to GMCs \citep{larson:gmc.scalings}. However, whether this can be responsible for reducing $\epsilon_{ff}$ is unclear, as \citet{federrath:2012.sfr.vs.model.turb.boxes} do not find particularly low $\epsilon_{ff}$ in their $\mathcal{M}=3$ simulations. Other alternatives would include some feedback mechanism that we have not accounted for, such as protostellar heating or outflows, or a systematic overestimation of inferred density of HCN clumps \citep{goldsmith:2017.electron.excitation}.


Caution is needed comparing the predicted cloud lifetimes to observationally inferred lifetimes, because this is sensitive both to the observational methods/tracers, and to the actual properties (e.g. mean densities) of the initial clouds (which we have freely varied, rather than drawing from a statistically representative sample of observed clouds). A detailed comparison will be the subject of future work (Grudi\'{c} et al. 2018, in prep). However, we can make some preliminary comparisons. Lee et al. 2016 estimate a mean GMC lifetime of $\sim \unit[24]{Myr}$ for a population of clouds with a median free-fall time of $\unit[6.7]{Myr}$ (corresponding to a mean density of $25$ $H_2$ molecules per $\unit[]{cm^{3}}$). Our $\Sigma=\unit[127]{M_\odot\,pc^{-2}}$, $R=\unit[50]{pc}$ run is the closest to this in mean density ($\unit[33]{cm^{-3}}$) and free-fall time, and its major star formation episode lasts for $2.5\,t_{ff,0} \approx \unit[15]{Myr}$ (See Figure \ref{eff_vs_t} and Table \ref{simlist}). This is somewhat smaller than observed, although similar enough that differences in how ``lifetime'' is measured and observationally estimated might account for the difference. Moreover, real GMCs are not, of course, isolated, but can accrete continuously over their lifetime and may have turbulence ``stirred'' externally which further can slow collapse (for a review, see \citealt{2010ARA&A..48..547F}). It seems likely, therefore, that clouds embedded in a realistic ISM would have somewhat longer lifetimes.


\subsection{Slow star formation}
The scaling and saturation of of $\epsilon_{ff}$ appears at first to be at odds with the notion of ``slow'' star formation, wherein it has been observed that $\epsilon_{ff} \sim 1 \%$ universally, from Milky Way-like to ULIRG-like environments \citep{kennicutt98, krumholz.schmidt,krumholz:sf.eff.in.clouds,krumholz:2012.universal.sf.efficiency}. This slow speed of star formation has been explained theoretically in terms of the properties of the turbulent ISM alone \citep[e.g.][]{krumholz.schmidt,hennebelle:2011.time.dept.imf.eps}, so it is necessary to compare the predictions of these theories with those of feedback-moderated star formation to determine whether feedback is a necessary part of the picture. In making this comparison, we emphasize that our prediction pertains to individual unstable clouds near virial equilibrium, and not to any significant patch of a galaxy that may contain GMCs in various states of formation and disruption, as well as the other phases of the ISM. In the latter case, it has been shown in \citet{hopkins:2013.fire} and \citet{orr:2016.what.fires.up.SF} that the same physical models used in our simulations also robustly predict that $\epsilon_{ff,gal}\sim 1\%$ on galactic scales on average, despite assuming that $\epsilon_{ff} = 1$ on the smallest resolvable scales, as star formation reaches a statistical equilibrium when smoothed on $\unit[>1]{kpc}$ scales.

Both the feedback-disrupted cloud picture suggested by our simulations and purely turbulence-regulated star formation theories successfully predict the median value $\epsilon_{ff} \sim 1\%$ in Milky Way GMCs, however they do so for completely different physical reasons. However, the observed {\it dispersion} in $\epsilon_{ff}$ for a given set of cloud conditions has not been found to be less than \unit[0.5]{dex} \citep{heiderman:2010.gmcs,evans:2014.sfe,heyer:2016.clumps,lee:2016.gmc.eff,vuti:2016.gmcs}; \citeauthor{lee:2016.gmc.eff} found $\unit[0.91]{dex}$ from the most complete Milky Way GMC dataset that we are aware of. As they noted, the turbulence-regulated models do not predict this much scatter because they do not allow for $\epsilon_{ff}$ to vary for a given set of turbulent ISM conditions. \citeauthor{lee:2016.gmc.eff} showed that the scatter can arise from observing GMCs at random points in their lifetime of initial collapse, star formation, and feedback disruption. For Milky Way-like conditions, our simulations do predict intrinsic dispersions in $\epsilon_{ff}$ of the same order as what has been observed; whether the figure of $\unit[0.91]{dex}$ can be fully accounted for depends upon the relationship between $\epsilon_{ff}$ and its observational proxy, which we will address in future work.

The gas-rich nuclei in Arp 220 provide an interesting case study for the speed of star formation. The total SFR of $\unit[240]{\msun\,yr^{-1}}$, inferred from its IR luminosity \citep{downes.solomon:ulirgs,kennicutt:1998.review}, appears to agree nicely with the theory of slow star formation, yet our simulations at comparable gas surface density $\sim \unit[10^4]{\msun\,pc^{-2}}$ predict $\epsilon_{ff}\sim 20\%$. Considering several $\unit[10^9]{\msun}$ of gas localized within two disks, each with radius smaller than $\unit[100]{pc}$ \citep{scoville:2016.arp220}, the resulting SFR should be well in excess of $\unit[10^3]{\msun\,yr^{-1}}$, an order of magnitude greater than the $L_{IR}$-inferred value. Our simulations do not consider the stabilization of the gas disk due to the presence of the central SMBH, but this can probably only reduce the predicted SFR by a factor of a few \citep{utreras:2016.rotation}. The apparent discrepancy may lie in the use of $L_{IR}$ to determine the SFR, as it only provides an average value over the lifetime of OB stars, $\unit[4]{Myr}$. Because the dynamical time in the nuclear disks is of order $\unit[10^5]{yr}$ \citep{scoville:2016.arp220}, it is unlikely that the SFR has been steady over this comparatively long averaging window. Estimates of the SFR from supernova rates have the same limitation. Therefore, the possibility that the SFR in Arp 220 has recently been in excess of $\unit[10^3]{\msun\,yr^{-1}}$ cannot be excluded on this basis \citep{arp.220.fast.sf,arp.220.fast.sf2}.

\subsection{Comparison with other GMC star formation studies}

Many numerical studies have been performed that are conceptually similar to the ones in this paper, following the collapse of an idealized turbulent cloud and the resulting star formation and feedback processes. It is useful to compare and contrast our predictions with these studies, in particular in cases where specific feedback mechanisms have been considered in greater detail.

Our run without stellar feedback is most comparable with previous simulations of isothermal supersonic MHD turbulence with gravity \citep{kritsuk:2011.density.pdf.power.law, collins:2012.mhd.sf,padoan:2012.sfe,lee:2015.gravoturbulence}. In these simulations, the SFR tends to grow until $\epsilon_{ff}$ is of order unity, with its particular value depending somewhat upon the regular and Alfv\'{e}nic Mach numbers, the virial parameter, and the details of the turbulent driving, and the final $\epsilon_{int}\sim1$ due to the lack of feedback. The value $\epsilon_{ff}=0.52$ we obtain in feedback-free cloud collapse without feedback is most consistent with the \citet{federrath:2012.sfr.vs.model.turb.boxes} models with mixed or solenoidal driving.

\citet{dale:2012} ran a parameter study of feedback-disrupted clouds, considering only photoionization heating. We have found in tests that photoionization heating only is insufficient to disrupt a cloud with an escape velocity that is large compared to the sound speed $c_s \sim \unit[10]{km\,s^{-1}}$ of photoionized gas. This agrees with the trend of \citet{dale:2012}, which found order-unity $\epsilon_{int}$ in clouds with high escape velocity (Runs `X' and `F'). Also, our $M=\unit[10^4]{\msun\,pc^{-2}}$, $R=\unit[5]{pc}$ has the same physical parameters as Run `J' in \citet{dale:2012}. This had $\epsilon_{int}=0.04$, while the final stellar mass in Run `J' was $35\%$ and rising at $\unit[3.5]{Myr}$. We re-simulated this run with photoionization heating only and radiation pressure only, and the one with photoionization heating had a very similar star formation history and cloud morphology to Run `J'. The one with radiation pressure only had $\epsilon_{int}=0.05$, very close to the full physics run. Radiation pressure is thus the primary feedback mechanism even in this region of parameter space where photoionization heating alone could still theoretically disrupt the cloud.

The radiation hydrodynamics star formation simulations of \citet{raskutti:2016.gmcs} focus upon the effects of stellar feedback from the single-scattered monochromatic photons at a high opacity corresponding to UV photons. They use the radiation hydrodynamics code {\small Hyperion}, evolving the radiation field on a fixed grid according to the M1 closure \citep{skinner:2013.hyperion}. They overpredict the efficiency of their fiducial Milky Way-like GMC run by an order of magnitude, obtaining $\epsilon_{int} = 0.43$ for a cloud with $M=\unit[5 \times 10^4]{\msun}$ and $R=\unit[15]{pc}$, which has average surface density $\unit[70]{\msun\,pc^{-2}}$. Extrapolating our simulation results using equation \ref{SFEfit} gives $\epsilon_{int}=0.02$ for a cloud with these parameters, in much better agreement with observations (Section \ref{observations} and references therein). We have found that $\epsilon_{int}\sim 0.04$ in a test run with otherwise similar initial conditions to \citeauthor{raskutti:2016.gmcs} and radiation pressure as the only feedback (Appendix \ref{rptest}). 

This order of magnitude discrepancy may be due to the behaviour of the M1 closure in such an optically-thick, multi-source radiative transfer problem. Experiments in developing {\small GIZMO}'s own M1 RHD scheme have shown that the momentum imparted to the gas by the radiation field around an embedded source can be underestimated by an order of magnitude if the attenuation length $\lambda = \rho^{-1}\kappa_{UV}^{-1}$ is not well-resolved, which it certainly is not at the densities, opacities, and spatial resolution typical in the \citeauthor{raskutti:2016.gmcs} simulations \footnote{This problem is averted by the shell-driving test problem presented by \citeauthor{raskutti:2016.gmcs}, because the radiation first propagates through an optically-thin medium where the field is well-resolved.}. Secondly, photons propagated via the M1 scheme behave collisionally: colliding streams will form a shock rather than passing through each other. As stars form in a tightly-clustered configuration in isothermal fragmentation \citep{guszejnov:2016.correlation.function,guszejnov:2017.toy.model}, neighbouring stars  particles can cancel each other's fluxes. In summary, it is reasonable to suspect that ability of radiation pressure to disrupt the GMC was underestimated.

\citet{tsang:2017.ssc.rp} simulated super star cluster formation in cloud with mass $\unit[10^7]{\msun}$ and diameter $\unit[25]{pc}$, for a mean surface density of $\unit[1.6\times 10^4]{\msun\,pc^{-2}}$, comparable to the densest runs in our parameter study. They accounted for feedback via infrared radiation pressure, which is expected to dominate, with an accelerated Monte Carlo scheme that is more realistic than our more approximate treatment. They found that radiation pressure reduced $\epsilon_{int}$ by $\sim 30\%$ compared to the run with no feedback. Our simulations at this surface density had $\epsilon_{int}\sim0.64$, compared to $0.86+$ with no feedback, so despite our different treatments of radiation pressure the agreement is quite good.

It should be noted that most star formation in all simulations mentioned in this subsection occurs within some small ($\sim 2-3$) number of global freefall times, regardless of the final $\epsilon_{int}$ if the cloud is disrupted. This naturally leads to the linear relation between $\epsilon_{int}$ and $\epsilon_{ff}$ shown in Section \ref{section:results:eff}, suggesting that this is a very general feature of the star formation-cloud disruption process, insensitive to the details of stellar feedback. The role of feedback on cloud scales is to make star formation less efficient in a given amount of time, not to prolong the star-forming lifetime as it does on galactic scales.

\subsection{Bound star cluster formation}

$\epsilon_{int}$ should be an important quantity for the formation of bound star clusters. If all other factors are equal, the fraction of a star cluster remaining gravitationally bound after gas expulsion should increase with $\epsilon_{int}$ \citep{tutukov:1978,hills:1980,mathieu:1983,lada:1984,1985ApJ...294..523E}\footnote{Other factors influencing the bound fraction of a cluster include the virial state of the stars at gas expulsion \citet{goodwin:2009.cluster.formation} and the degree of initial degree of clumpy substructure \citep{smith:2011.cluster.assembly,smith:2013.cluster.assembly}}. It can thus be argued that the bound cluster formation efficiency $\Gamma$, the fraction of stars found in bound clusters, is a function of $\epsilon_{int}$, and hence of $\Sigma_{tot,0}$ by equation \ref{SFEformula}.  If equation \ref{SFEformula} holds, then cluster formation should be generic to regions of high $\Sigma_{gas}$. And indeed, rich populations of young bound clusters are ubiquitous in dense nuclear starbursts, including notable examples Arp 220 \citep{wilson:2006.arp220.superclumps}, M82 \citep{mccrady:m82.sscs}, and M83 \citep{bastian:2012.m83.clusters,ryon:2015.m83.clusters}. However, whether there actually is a general scaling in $\Gamma$ that depends on a single environmental parameter associated with surface density is currently an open problem: \citet{adamo:2015.cfe} and \citet{johnson:2016.cluster.formation.efficiency} appear to support this hypothesis, while \citet{chandar:2015.cfe} does not. 

GMCs in the Milky Way and other nearby galaxies typically have $\Sigma_{gas} \sim \unit[100]{\msun pc^{-2}}$ \citep{larson:gmc.scalings,solomon:gmc.scalings}, giving $\epsilon_{int}\sim 3\%$ at best, yet young bound star clusters are still observed to have formed within the galaxy \citep{portegies-zwart:2010.starcluster.review}. Rather than simply turning off below a certain surface density threshold, $\Gamma$ is theoretically expected to scale smoothly as a function of $\Sigma_{gas}$, saturating to a value of $\sim 70\%$ \citep{kruijssen:2012.cluster.formation.efficiency}. Star cluster formation may be possible in environments that are less dense on average because star-forming clouds are hierarchically structured, with a broad surface density PDF. If $\epsilon_{int}$ is determined in a scale-free fashion according to equation \ref{SFEformula}, it will apply just as well on the scale of denser-than-average subclumps once they decouple from their environment, allowing them to have high $\epsilon_{int}$ locally even if $\epsilon_{int}$ is small on larger scales (e.g. \citet{2012MNRAS.419..841K}). If this argument is valid, we expect to see some amount of bound cluster formation in any star-forming environment.

The production of bound star clusters is generally associated with high-pressure environments, where the pressure associated with the midplane of a galactic disk can be estimated as $P \sim G \Sigma_{gas} \Sigma_{tot}$ \citep{elmegreen:1997.open.closed.cluster.same.mf.form}. \citeauthor{elmegreen:1997.open.closed.cluster.same.mf.form} proposed a picture wherein GMCs are confined by this pressure $P\sim \rho v_t^2$, rather than their self-gravity, and the gas mass loss rate in a protocluster was assumed to be $\dot{M} \propto L/v_t^2$, where $L$ is the protocluster luminosity. Thus the fraction of the gas mass converted to stars with fixed $\epsilon_{ff}$ is greater when $P$ is greater. The picture suggested by the simulations in this paper is presents an alternative to this;  $\epsilon_{ff}$ is not fixed, and the timescale of mass loss is always on the order of the freefall time. Clouds are confined by self-gravity, rather than external pressure, and their SFE is greater at greater $P\sim G \Sigma_{gas}^2$ because of the relative scaling of the strength of feedback and self-gravity.



In future work we will use these simulations to study the mapping between galactic environments and the populations of bound star clusters they produce, providing the stepping stone between lower-resolution cosmological simulations and single-cluster dynamical studies. This development is necessary, in particular, for the theory of cosmological SMBH seed formation from runaway stellar mergers in dense clusters (see \citet{portegies-zwart:2002, mouri:2002.runaway.black.holes, gurkan2004,devecchi:2009.smbh.clusters}). It would also allow a more self-consistent model of pairing and evolution of the population of massive ($\sim 60\msun$) black hole binaries like the progenitor of GW150914 \citep{GW150914}; a significant fraction of these are expected to be found in bound star clusters \citep{rodriguez:2015.bbh.globulars, rodriguez:2016.bbh.globulars}.

\subsection{The nature of nuclear star formation}

Our results here illustrate the claim of \citet{torrey:2016.feedback.instability}: {\it no equilibrium exists} for gas-rich nuclear disks with short dynamical times, and their dynamics have an inherently transient nature: they undergo rapid fragmentation followed by rapid gas expulsion. Star-forming nuclear disk calculations {\it must} account for stellar feedback in a way that is appropriate to their short time-scales, or else risk obtaining unphysical solutions. This caveat may very well limit the validity of isolated nuclear disk simulations that use a \citet{springel:multiphase}-like effective-EOS ISM model and a slow sub-grid star formation law, both of which have been widely used in the field of galaxy simulations. For example, \citet{hopkins:zoom.sims} simulated circumnuclear disks of similar mass and radius to the ones in this paper, but in absence of the appropriate feedback physics the SFR of the disks was quite likely underestimated by at least an order of magnitude.

A robust result of our simulations is that both $\epsilon_{int}$ and $\epsilon_{ff}$ must saturate to $\sim 1$ at surface densities in excess of $\unit[10^4]{\msun pc^{-2}}$. Barring other unaccounted-for feedback physics (see Section \ref{caveats}), and neglecting environmental interactions, we conclude that a gas-dominated cloud with $\Sigma_{gas}>>\unit[10^3]{\msun\,pc^{-2}}$ will convert nearly all of its gas to stars in a few crossing times. In this limit, we expect a result similar to our simulations: a population of massive star clusters will form, and will eventually merge into a single cluster because the high global SFE will allow the system to remain bound. If a relatively low-mass SMBH is present, it may sink to the centre of this cluster under dynamical friction. However, it is also possible that before the final nuclear cluster has formed, the SMBH and clusters effectively behave as a few-$N$-body system, which has chaotic behaviour and often results in the ejection of one or more members. Such ejections will prolong the time necessary for SMBH to form binary pairs in galaxy mergers, and may lower the resulting low-frequency gravitational wave background.

If star formation occurs near an SMBH, the gravity of the SMBH also contributes to the binding force on the gas. If we re-derive \ref{SFEformula} and consider only the force of gravity of the SMBH on the gas, we obtain a lower bound for the integrated SFE of a gas disk of radius $R$ around a black hole of mass $M_{BH}$:
\be
\epsilon_{int} \geq \left(1 + \frac{\mathit{\pi} R^2 \Sigma_{crit}}{M_{BH}} \right)^{-1}.
\ee
This assumes that the gas is not somehow being prevented from forming stars by AGN feedback and that the dynamical effect of the black hole upon the gas flow does not slow star formation enough to make the gas consumption time longer than $\sim\unit[10]{Myr}$. The characteristic radius at which $\epsilon_{int}$ saturates to $\sim1$ is then:
\be
R_{SF} \sim \sqrt{M_{BH} /2 \pi \Sigma_{crit}} = \unit[6]{pc} \left( \frac{M_{BH}}{10^6 \msun} \right)^\frac{1}{2},
\ee
using $\Sigma_{crit}=\unit[2800]{\msun \, pc^{-2}}$. 

Under these assumptions, the in-situ formation of a nuclear star cluster could proceed as follows: if enough low-angular momentum gas falls within $R_{SF}$ of an SMBH to become gravitationally unstable, it will be rapidly consumed by star formation, leaving behind a nuclear star cluster and little remaining gas. The fiducial value $\unit[6]{pc}$ derived here does lie in the range of effective radii of nuclear star clusters found in several different types of galaxies (see \citet{hopkins:maximum.surface.densities} and references therein).

Such efficient star formation near black holes may have drastic implications for the ability of gas from the galactic disk to be accreted onto a central SMBH, as the gas may fragment into stars before reaching the hole within a few dynamical times, at which point it can no longer lose angular momentum efficiently. This contrasts greatly with models which assume star formation must be slow ($\epsilon_{ff} \sim 1\%$) all the way down to the black hole; in this case, a steady supply of gas can reach the black hole even with modest torques, as gas has $\sim 100$ dynamical times to lose its angular momentum before being converted to stars. As such, it is important that studies of AGN accretion on $\unit[\sim]{pc}$ and smaller scales consider the physics of the multiphase ISM and star formation in some detail.

\subsection{Absence of metal-enriched supermassive direct-collapse objects}
These simulations were originally conceived as an attempt to reproduce the mechanism for direct-collapse supermassive black hole formation simulated in \citet{mayer:2009.direct.collapse.bh} and \citet{mayer:2015.direct.collapse.bh} with a more realistic approach to cooling and star formation. To summarize, these works propose that in the gas-rich nuclear disk resulting from a galaxy merger, fragmentation can be suppressed by some combination of turbulence and suppression of cooling due to optical thickness, enabling accretion onto a supermassive quasi-star even for ISM with solar metal abundances. To avoid over-cooling in optically thick regions, we implemented the optically-thick cooling approximation of \citet{rafikov:2007.convect.cooling.grav.instab.planets} so as to interpolate between the optically-thin and -thick cooling regimes where appropriate. In previous tests we also chose a rather high ($10^7 cm^{-3}$) density threshold for star formation and allowed star formation only when the local Jeans mass is $<10^3\msun$, so as to prevent premature conversion of gas particles into star particles where they may otherwise form a supermassive object. Our simulations reach comparable optical depths and turbulent velocity dispersions to the nuclear disks in the Mayer simulations, however we report no formation of direct-collapse objects. In numerical experiments, we have only been able to produce anything resembling a supermassive quasi-star if we implement a temperature floor of $\unit[10^4]{K}$ and slow the local star formation rate $\dot{\rho}_\star$ to $\unit[1]{\%}$ of the usual value. As these are similar to the choices made for \citet{mayer:2009.direct.collapse.bh} and \citet{mayer:2015.direct.collapse.bh}, it seems that metal-enriched direct-collapse object formation is a numerical artifact of slow subgrid star formation and a lack of low-temperature cooling. Our conclusions agree with those obtained by \citet{ferrara:2013.smbh.seeds} using a one-dimensional disk model: if realistic low-temperature cooling is accounted for, the cooling time in the metal-enriched ISM is invariably too short to suppress fragmentation down to the scales required to directly form a supermassive object.

\subsection{Feedback physics uncertainties}
\label{caveats}
Most of what is known about the effects of stellar feedback on GMC scales has been learned from observations of star-forming complexes within the Milky Way, and even then the true efficiencies of many feedback mechanisms acting in Milky Way-like environments are still loosely constrained, to say nothing of generalizing these mechanisms to ULIRG-like environments. Here we list uncertainties in the strength of feedback which could conceivably affect our results:
\subsubsection{The Initial Mass Function}
Throughout this work, we have assumed that the initial stellar mass function, and hence $\nicefrac{\dot{P}_\star}{m_\star}$, is independent of the environment of star formation. If the IMF were to become more top-heavy in environments of high surface density,  $\nicefrac{\dot{p}_\star}{m_\star}$ would increase, and as our simulations have shown, this is the quantity to which our results are most sensitive. Supposing that $\frac{\dot{P}_\star}{m_\star}$ did scale at least linearly with $\Sigma_{gas}$ due to enhanced type O star production, this would limit the maximum star formation efficiency. There is some observational evidence of a dearth of low-mass stars in dense nuclear environments \citep{smith:2001.m82.ssc, bartko:2010.mw.nucdisk.imf}, however such observations can be subject to significant sampling bias because the time-scale for mass segregation is short in dense systems. For this reason and others, \citet{bastian:2010.imf.universality} concluded that current observations were still largely consistent with a universal IMF.

\subsubsection{Infrared radiation pressure}
Radiation pressure plays an important role in the feedback budget in many of our simulations; even in cases where the final gas blowout is ultimately due to SNe, radiation helps prevent an initial runaway of the SFE before SNe start to occur. We have found that $\epsilon_{int}$ saturates to a value close to 1 as surface density becomes large, however \citet{murray:molcloud.disrupt.by.rad.pressure} argued that the IR opacity of dust grains should limit the saturation point of $\epsilon_{int}$ for gas with solar abundances, as radiation pressure in the optically thick regime is the only force of feedback which can conceivably scale as fast as the gas self-gravity. By this argument, the saturation SFE $\epsilon^{max}_{int}$ is expected to scale $\sim \left(\kappa_{IR} \Sigma_{crit}\right)^{-1}$, which takes a value of $\sim \frac{1}{2}$ for gas with solar metal abundances. However, in a realistic, 3-dimensional scenario where hydrodynamics is coupled to the radiation field in an inhomogeneous ISM, it is actually unlikely that radiation pressure can achieve the whole ``$\tau_{IR}$ boost'', either because photons will have a tendency to leak out of the most optically-thin lines of sight, or because the radiative Rayleigh-Taylor instability is able to efficiently dissipate kinetic energy \citep{krumholz:2012.rrt}. Radiation hydrodynamics studies on this problem are ongoing (see also \citet{krumholz:2013.rhd,rosdahl:2015.rhd,tsang:2015.rhd, davis:ulirg.rhd,skinner:2015.ir.molcloud.disrupt,zhang:2017.ir.pressure}), and although results have varied with the radiative transfer scheme used, they do generally agree that the scaling of the momentum deposited to the gas with the mean $\tau_{IR}$ is sublinear for sufficiently large $\tau_{IR}$, forcing the integrated SFE to ultimately saturate to $\sim 1$.

\section{Summary}
We have performed a parameter study of 3D multi-physics MHD simulations of star-forming gas disks with initial parameters spanning two orders of magnitude in surface density and in spatial scale, including the physics of supernovae, stellar winds, radiation pressure, and photoionization heating. Due to the generality of the simulation setup, we have been able to study the nature of star formation in gas-rich environments in general, including nuclear starbursts and GMCs. Our main findings are as follows:
\begin{itemize}
\item In any bound, gas-rich star-forming cloud with short ($\sim \mathrm{10Myr}$ or less) dynamical time, star formation proceeds until it causes an inevitable gas blowout, with the final SFE determined mainly by the balance of feedback and gravitation, with other physical mechanisms having secondary importance. 
\item The integrated SFE $\epsilon_{int}$ of such a system scales strongly with the initial gas surface density $\Sigma_{tot,0}$ with weak dependence upon other parameters, and saturates to a value $\sim 1$ at adequately high surface density, despite the effects of strong feedback. We find good agreement with analytic derivations of $\epsilon_{int}$ which take the form of equation \ref{SFEformula} \citep{fall:2010.sf.eff.vs.surfacedensity, murray:molcloud.disrupt.by.rad.pressure, dekel:2013.giant.clumps, thompson:2016.eddington.outflows}, fitting a value $\Sigma_{crit}=\unit[2800]{\msun\,pc^{-2}}$ from the simulations. The agreement across different spatial scales is non-trivial and somewhat surprising, as our parameter space bridges distinct time-scale regions where radiation pressure ($<\unit[3]{Myr}$) and SN explosions ($>\unit[3]{Myr}$) dominate the feedback energy and momentum budget. The prediction of this SFE model is that $\epsilon_{int}$ in self-gravitating clouds should scale from $\sim1\%$ at $\unit[10^2]{\msun\,pc^{-2}}$ and $\sim 10\%$ at $\unit[10^3]{\msun\,pc^{-2}}$, as is found in local GMCs and dense clumps (Section \ref{observations} and references therein). The model also predicts that SFE ultimately saturates to $\sim 100\%$ in the limit of very high surface density.



\item We find a proportional relation between the integrated SFE $\epsilon_{int}$ and the per-freefall SFE $\epsilon_{ff}$ (equation \ref{eff_vs_eint}) for self-gravitating clouds, essentially because the clouds always produce enough stars to self-destruct within $\sim 2-3$ dynamical times. $\epsilon_{ff}$ is determined only initially by such details as cooling and magnetic fields, and will inevitably grow until moderated by stellar feedback. The observed $\epsilon_{ff}$ distribution for Milky Way GMCs can be accounted for by combining the spread from this relation and a modest intrinsic spread due to the time-varying SFE of a single cloud. The variation in $\epsilon_{ff}$ is at odds with a universal slow star formation ($\epsilon_{ff}\sim1\%$) law when applied to individual clouds, but the same physics used in this study recover the $\epsilon_{ff,gal}\sim1\%$ relation in cosmological simulations \citep{hopkins:2013.fire,fire2}.

Thus we have determined the basic properties of feedback-moderated star formation for self-gravitating, unstable gas complexes. In a subsequent paper, we have used these simulations to study the process of star cluster assembly \citep{grudic:2017}. Future work will elucidate the relation between theoretical predictions of cloud SFE and its observational proxies, the mapping between galactic environmental properties and populations of star clusters, and the detailed dynamical history of star cluster formation as determined by feedback.

\end{itemize}
\acknowledgments
We thank Neal J. Evans II, Eve Ostriker, D\'avid Guszejnov, Chris Hayward, Matthew Orr and Andrew Wetzel for helpful comments and critique. We also thank the anonymous referees for highly comprehensive and helpful feedback that motivated a more thorough understanding of our results. Support for PFH and MG was provided by an Alfred P. Sloan Research Fellowship, NASA ATP Grant NNX14AH35G and NSF Collaborative Research Grant $\#1411920$ and CAREER grant $\#1455342$. Numerical calculations were run on the Caltech computer cluster `Zwicky' (NSF MRI award $\#$PHY-0960291) and allocation TG-AST130039 granted by the Extreme Science and Engineering Discovery Environment (XSEDE) supported by the NSF.
\vspace{0.2cm}

\bibliographystyle{mnras}
\bibliography{master.bib}

\begin{thebibliography}{}
\makeatletter
\relax
\def\mn@urlcharsother{\let\do\@makeother \do\$\do\&\do\#\do\^\do\_\do\%\do\~}
\def\mn@doi{\begingroup\mn@urlcharsother \@ifnextchar [ {\mn@doi@}
  {\mn@doi@[]}}
\def\mn@doi@[#1]#2{\def\@tempa{#1}\ifx\@tempa\@empty \href
  {http://dx.doi.org/#2} {doi:#2}\else \href {http://dx.doi.org/#2} {#1}\fi
  \endgroup}
\def\mn@eprint#1#2{\mn@eprint@#1:#2::\@nil}
\def\mn@eprint@arXiv#1{\href {http://arxiv.org/abs/#1} {{\tt arXiv:#1}}}
\def\mn@eprint@dblp#1{\href {http://dblp.uni-trier.de/rec/bibtex/#1.xml}
  {dblp:#1}}
\def\mn@eprint@#1:#2:#3:#4\@nil{\def\@tempa {#1}\def\@tempb {#2}\def\@tempc
  {#3}\ifx \@tempc \@empty \let \@tempc \@tempb \let \@tempb \@tempa \fi \ifx
  \@tempb \@empty \def\@tempb {arXiv}\fi \@ifundefined
  {mn@eprint@\@tempb}{\@tempb:\@tempc}{\expandafter \expandafter \csname
  mn@eprint@\@tempb\endcsname \expandafter{\@tempc}}}

\bibitem[\protect\citeauthoryear{{Abbott} et~al.,}{{Abbott}
  et~al.}{2016}]{GW150914}
{Abbott} B.~P.,  et~al., 2016, \mn@doi [Physical Review Letters]
  {10.1103/PhysRevLett.116.061102}, \href
  {http://adsabs.harvard.edu/abs/2016PhRvL.116f1102A} {116, 061102}

\bibitem[\protect\citeauthoryear{{Adamo}, {Kruijssen}, {Bastian}, {Silva-Villa}
   \& {Ryon}}{{Adamo} et~al.}{2015}]{adamo:2015.cfe}
{Adamo} A.,  {Kruijssen} J.~M.~D.,  {Bastian} N.,  {Silva-Villa} E.,   {Ryon}
  J.,  2015, \mn@doi [\mnras] {10.1093/mnras/stv1203}, \href
  {http://adsabs.harvard.edu/abs/2015MNRAS.452..246A} {452, 246}

\bibitem[\protect\citeauthoryear{{Adamo} et~al.,}{{Adamo}
  et~al.}{2017}]{adamo:2017.cfe}
{Adamo} A.,  et~al., 2017, \mn@doi [\apj] {10.3847/1538-4357/aa7132}, \href
  {http://adsabs.harvard.edu/abs/2017ApJ...841..131A} {841, 131}

\bibitem[\protect\citeauthoryear{{Agertz}, {Kravtsov}, {Leitner}  \&
  {Gnedin}}{{Agertz} et~al.}{2013}]{agertz:2013.new.stellar.fb.model}
{Agertz} O.,  {Kravtsov} A.~V.,  {Leitner} S.~N.,   {Gnedin} N.~Y.,  2013,
  \mn@doi [\apj] {10.1088/0004-637X/770/1/25}, \href
  {http://adsabs.harvard.edu/abs/2013ApJ...770...25A} {770, 25}

\bibitem[\protect\citeauthoryear{{Anantharamaiah}, {Viallefond}, {Mohan},
  {Goss}  \& {Zhao}}{{Anantharamaiah} et~al.}{2000}]{arp.220.fast.sf}
{Anantharamaiah} K.~R.,  {Viallefond} F.,  {Mohan} N.~R.,  {Goss} W.~M.,
  {Zhao} J.~H.,  2000, \mn@doi [\apj] {10.1086/309063}, \href
  {http://adsabs.harvard.edu/abs/2000ApJ...537..613A} {537, 613}

\bibitem[\protect\citeauthoryear{{Bartko} et~al.}{{Bartko}
  et~al.}{2010}]{bartko:2010.mw.nucdisk.imf}
{Bartko} H.,  et~al., 2010, \mn@doi [\apj] {10.1088/0004-637X/708/1/834}, \href
  {http://adsabs.harvard.edu/abs/2010ApJ...708..834B} {708, 834}

\bibitem[\protect\citeauthoryear{{Bastian}, {Covey}  \& {Meyer}}{{Bastian}
  et~al.}{2010}]{bastian:2010.imf.universality}
{Bastian} N.,  {Covey} K.~R.,   {Meyer} M.~R.,  2010, \mn@doi [\araa]
  {10.1146/annurev-astro-082708-101642}, \href
  {http://adsabs.harvard.edu/abs/2010ARA%26A..48..339B} {48, 339}

\bibitem[\protect\citeauthoryear{{Bastian} et~al.,}{{Bastian}
  et~al.}{2012}]{bastian:2012.m83.clusters}
{Bastian} N.,  et~al., 2012, \mn@doi [\mnras]
  {10.1111/j.1365-2966.2011.19909.x}, \href
  {http://adsabs.harvard.edu/abs/2012MNRAS.419.2606B} {419, 2606}

\bibitem[\protect\citeauthoryear{{Bate}, {Bonnell}  \& {Price}}{{Bate}
  et~al.}{1995}]{bate:1995.protobinary.accretion.vs.frag}
{Bate} M.~R.,  {Bonnell} I.~A.,   {Price} N.~M.,  1995, \mnras, \href
  {http://adsabs.harvard.edu/abs/1995MNRAS.277..362B} {277, 362}

\bibitem[\protect\citeauthoryear{{Bolatto}, {Leroy}, {Rosolowsky}, {Walter}  \&
  {Blitz}}{{Bolatto} et~al.}{2008}]{bolatto:2008.gmc.properties}
{Bolatto} A.~D.,  {Leroy} A.~K.,  {Rosolowsky} E.,  {Walter} F.,   {Blitz} L.,
  2008, \mn@doi [\apj] {10.1086/591513}, \href
  {http://adsabs.harvard.edu/abs/2008ApJ...686..948B} {686, 948}

\bibitem[\protect\citeauthoryear{{Bryant} \& {Scoville}}{{Bryant} \&
  {Scoville}}{1999}]{bryant.scoville:ulirgs.co}
{Bryant} P.~M.,  {Scoville} N.~Z.,  1999, \mn@doi [\aj] {10.1086/300879}, \href
  {http://adsabs.harvard.edu/abs/1999AJ....117.2632B} {117, 2632}

\bibitem[\protect\citeauthoryear{{Chandar}, {Fall}  \& {Whitmore}}{{Chandar}
  et~al.}{2015}]{chandar:2015.cfe}
{Chandar} R.,  {Fall} S.~M.,   {Whitmore} B.~C.,  2015, \mn@doi [\apj]
  {10.1088/0004-637X/810/1/1}, \href
  {http://adsabs.harvard.edu/abs/2015ApJ...810....1C} {810, 1}

\bibitem[\protect\citeauthoryear{{Collins}, {Kritsuk}, {Padoan}, {Li}, {Xu},
  {Ustyugov}  \& {Norman}}{{Collins} et~al.}{2012}]{collins:2012.mhd.sf}
{Collins} D.~C.,  {Kritsuk} A.~G.,  {Padoan} P.,  {Li} H.,  {Xu} H.,
  {Ustyugov} S.~D.,   {Norman} M.~L.,  2012, \mn@doi [\apj]
  {10.1088/0004-637X/750/1/13}, \href
  {http://adsabs.harvard.edu/abs/2012ApJ...750...13C} {750, 13}

\bibitem[\protect\citeauthoryear{{Dale}, {Ercolano}  \& {Bonnell}}{{Dale}
  et~al.}{2012}]{dale:2012}
{Dale} J.~E.,  {Ercolano} B.,   {Bonnell} I.~A.,  2012, \mn@doi [\mnras]
  {10.1111/j.1365-2966.2012.21205.x}, \href
  {http://adsabs.harvard.edu/abs/2012MNRAS.424..377D} {424, 377}

\bibitem[\protect\citeauthoryear{{Davis}, {Jiang}, {Stone}  \&
  {Murray}}{{Davis} et~al.}{2014}]{davis:ulirg.rhd}
{Davis} S.~W.,  {Jiang} Y.-F.,  {Stone} J.~M.,   {Murray} N.,  2014, \mn@doi
  [\apj] {10.1088/0004-637X/796/2/107}, \href
  {http://adsabs.harvard.edu/abs/2014ApJ...796..107D} {796, 107}

\bibitem[\protect\citeauthoryear{{Dekel} \& {Krumholz}}{{Dekel} \&
  {Krumholz}}{2013}]{dekel:2013.giant.clumps}
{Dekel} A.,  {Krumholz} M.~R.,  2013, \mn@doi [\mnras] {10.1093/mnras/stt480},
  \href {http://adsabs.harvard.edu/abs/2013MNRAS.432..455D} {432, 455}

\bibitem[\protect\citeauthoryear{{Devecchi} \& {Volonteri}}{{Devecchi} \&
  {Volonteri}}{2009}]{devecchi:2009.smbh.clusters}
{Devecchi} B.,  {Volonteri} M.,  2009, \mn@doi [\apj]
  {10.1088/0004-637X/694/1/302}, \href
  {http://adsabs.harvard.edu/abs/2009ApJ...694..302D} {694, 302}

\bibitem[\protect\citeauthoryear{{Downes} \& {Solomon}}{{Downes} \&
  {Solomon}}{1998}]{downes.solomon:ulirgs}
{Downes} D.,  {Solomon} P.~M.,  1998, \mn@doi [\apj] {10.1086/306339}, \href
  {http://adsabs.harvard.edu/abs/1998ApJ...507..615D} {507, 615}

\bibitem[\protect\citeauthoryear{{Elmegreen}}{{Elmegreen}}{1983}]{elmegreen:1983}
{Elmegreen} B.~G.,  1983, \mn@doi [\mnras] {10.1093/mnras/203.4.1011}, \href
  {http://adsabs.harvard.edu/abs/1983MNRAS.203.1011E} {203, 1011}

\bibitem[\protect\citeauthoryear{{Elmegreen}}{{Elmegreen}}{2000}]{elmegreen:2000}
{Elmegreen} B.~G.,  2000, \mn@doi [\apj] {10.1086/308361}, \href
  {http://adsabs.harvard.edu/abs/2000ApJ...530..277E} {530, 277}

\bibitem[\protect\citeauthoryear{{Elmegreen}}{{Elmegreen}}{2007}]{elmegreen:2007}
{Elmegreen} B.~G.,  2007, \mn@doi [\apj] {10.1086/521327}, \href
  {http://adsabs.harvard.edu/abs/2007ApJ...668.1064E} {668, 1064}

\bibitem[\protect\citeauthoryear{{Elmegreen} \& {Clemens}}{{Elmegreen} \&
  {Clemens}}{1985}]{1985ApJ...294..523E}
{Elmegreen} B.~G.,  {Clemens} C.,  1985, \mn@doi [\apj] {10.1086/163320}, \href
  {http://adsabs.harvard.edu/abs/1985ApJ...294..523E} {294, 523}

\bibitem[\protect\citeauthoryear{{Elmegreen} \& {Efremov}}{{Elmegreen} \&
  {Efremov}}{1997}]{elmegreen:1997.open.closed.cluster.same.mf.form}
{Elmegreen} B.~G.,  {Efremov} Y.~N.,  1997, \mn@doi [\apj] {10.1086/303966},
  \href {http://adsabs.harvard.edu/abs/1997ApJ...480..235E} {480, 235}

\bibitem[\protect\citeauthoryear{{Evans} II et~al.,}{{Evans}
  et~al.}{2009}]{evans:2009.sfe}
{Evans} II N.~J.,  et~al., 2009, \mn@doi [\apjs] {10.1088/0067-0049/181/2/321},
  \href {http://adsabs.harvard.edu/abs/2009ApJS..181..321E} {181, 321}

\bibitem[\protect\citeauthoryear{{Evans}, {Heiderman}  \&
  {Vutisalchavakul}}{{Evans} et~al.}{2014}]{evans:2014.sfe}
{Evans} II N.~J.,  {Heiderman} A.,   {Vutisalchavakul} N.,  2014, \mn@doi
  [\apj] {10.1088/0004-637X/782/2/114}, \href
  {http://adsabs.harvard.edu/abs/2014ApJ...782..114E} {782, 114}

\bibitem[\protect\citeauthoryear{{Fall}, {Krumholz}  \& {Matzner}}{{Fall}
  et~al.}{2010}]{fall:2010.sf.eff.vs.surfacedensity}
{Fall} S.~M.,  {Krumholz} M.~R.,   {Matzner} C.~D.,  2010, \mn@doi [\apjl]
  {10.1088/2041-8205/710/2/L142}, \href
  {http://adsabs.harvard.edu/abs/2010ApJ...710L.142F} {710, L142}

\bibitem[\protect\citeauthoryear{{Faucher-Gigu{\`e}re}, {Quataert}  \&
  {Hopkins}}{{Faucher-Gigu{\`e}re} et~al.}{2013}]{cafg:sf.fb.reg.kslaw}
{Faucher-Gigu{\`e}re} C.-A.,  {Quataert} E.,   {Hopkins} P.~F.,  2013, \mn@doi
  [\mnras] {10.1093/mnras/stt866}, \href
  {http://adsabs.harvard.edu/abs/2013MNRAS.433.1970F} {433, 1970}

\bibitem[\protect\citeauthoryear{{Federrath} \& {Klessen}}{{Federrath} \&
  {Klessen}}{2012}]{federrath:2012.sfr.vs.model.turb.boxes}
{Federrath} C.,  {Klessen} R.~S.,  2012, \mn@doi [\apj]
  {10.1088/0004-637X/761/2/156}, \href
  {http://adsabs.harvard.edu/abs/2012arXiv1209.2856F} {761, 156}

\bibitem[\protect\citeauthoryear{{Federrath}, {Schober}, {Bovino}  \&
  {Schleicher}}{{Federrath} et~al.}{2014}]{federrath:supersonic.turb.dynamo}
{Federrath} C.,  {Schober} J.,  {Bovino} S.,   {Schleicher} D.~R.~G.,  2014,
  \mn@doi [\apjl] {10.1088/2041-8205/797/2/L19}, \href
  {http://adsabs.harvard.edu/abs/2014ApJ...797L..19F} {797, L19}

\bibitem[\protect\citeauthoryear{{Ferrara}, {Haardt}  \&
  {Salvaterra}}{{Ferrara} et~al.}{2013}]{ferrara:2013.smbh.seeds}
{Ferrara} A.,  {Haardt} F.,   {Salvaterra} R.,  2013, \mn@doi [\mnras]
  {10.1093/mnras/stt1350}, \href
  {http://adsabs.harvard.edu/abs/2013MNRAS.434.2600F} {434, 2600}

\bibitem[\protect\citeauthoryear{{Freeman}, {Rosolowsky}, {Kruijssen},
  {Bastian}  \& {Adamo}}{{Freeman} et~al.}{2017}]{freeman:2017.m83.gmcs}
{Freeman} P.,  {Rosolowsky} E.,  {Kruijssen} J.~M.~D.,  {Bastian} N.,   {Adamo}
  A.,  2017, \mn@doi [\mnras] {10.1093/mnras/stx499}, \href
  {http://adsabs.harvard.edu/abs/2017MNRAS.468.1769F} {468, 1769}

\bibitem[\protect\citeauthoryear{{Fukui} \& {Kawamura}}{{Fukui} \&
  {Kawamura}}{2010}]{2010ARA&A..48..547F}
{Fukui} Y.,  {Kawamura} A.,  2010, \mn@doi [\araa]
  {10.1146/annurev-astro-081309-130854}, \href
  {http://adsabs.harvard.edu/abs/2010ARA%26A..48..547F} {48, 547}

\bibitem[\protect\citeauthoryear{{Goldsmith} \& {Kauffmann}}{{Goldsmith} \&
  {Kauffmann}}{2017}]{goldsmith:2017.electron.excitation}
{Goldsmith} P.~F.,  {Kauffmann} J.,  2017, \mn@doi [\apj]
  {10.3847/1538-4357/aa6f12}, \href
  {http://adsabs.harvard.edu/abs/2017ApJ...841...25G} {841, 25}

\bibitem[\protect\citeauthoryear{{Goodwin}}{{Goodwin}}{2009}]{goodwin:2009.cluster.formation}
{Goodwin} S.~P.,  2009, \mn@doi [\apss] {10.1007/s10509-009-0116-5}, \href
  {http://adsabs.harvard.edu/abs/2009Ap%26SS.324..259G} {324, 259}

\bibitem[\protect\citeauthoryear{{Grudi{\'c}}, {Guszejnov}, {Hopkins},
  {Lamberts}, {Boylan-Kolchin}, {Murray}  \& {Schmitz}}{{Grudi{\'c}}
  et~al.}{2017}]{grudic:2017}
{Grudi{\'c}} M.~Y.,  {Guszejnov} D.,  {Hopkins} P.~F.,  {Lamberts} A.,
  {Boylan-Kolchin} M.,  {Murray} N.,   {Schmitz} D.,  2017, preprint, \href
  {http://adsabs.harvard.edu/abs/2017arXiv170809065G} {} (\mn@eprint {arXiv}
  {1708.09065})

\bibitem[\protect\citeauthoryear{{G{\"u}rkan}, {Freitag}  \&
  {Rasio}}{{G{\"u}rkan} et~al.}{2004}]{gurkan2004}
{G{\"u}rkan} M.~A.,  {Freitag} M.,   {Rasio} F.~A.,  2004, \mn@doi [\apj]
  {10.1086/381968}, \href {http://adsabs.harvard.edu/abs/2004ApJ...604..632G}
  {604, 632}

\bibitem[\protect\citeauthoryear{{Guszejnov}, {Hopkins}  \&
  {Krumholz}}{{Guszejnov} et~al.}{2016}]{guszejnov:2016.correlation.function}
{Guszejnov} D.,  {Hopkins} P.~F.,   {Krumholz} M.~R.,  2016, preprint, \href
  {http://adsabs.harvard.edu/abs/2016arXiv161000772G} {} (\mn@eprint {arXiv}
  {1610.00772})

\bibitem[\protect\citeauthoryear{{Guszejnov}, {Hopkins}  \&
  {Grudi{\'c}}}{{Guszejnov} et~al.}{2017}]{guszejnov:2017.toy.model}
{Guszejnov} D.,  {Hopkins} P.~F.,   {Grudi{\'c}} M.~Y.,  2017, preprint, \href
  {http://adsabs.harvard.edu/abs/2017arXiv170705799G} {} (\mn@eprint {arXiv}
  {1707.05799})

\bibitem[\protect\citeauthoryear{{Gutermuth}, {Megeath}, {Myers}, {Allen},
  {Pipher}  \& {Fazio}}{{Gutermuth} et~al.}{2009}]{gutermuth:2009.ysos}
{Gutermuth} R.~A.,  {Megeath} S.~T.,  {Myers} P.~C.,  {Allen} L.~E.,  {Pipher}
  J.~L.,   {Fazio} G.~G.,  2009, \mn@doi [\apjs] {10.1088/0067-0049/184/1/18},
  \href {http://adsabs.harvard.edu/abs/2009ApJS..184...18G} {184, 18}

\bibitem[\protect\citeauthoryear{{Heiderman}, {Evans}, {Allen}, {Huard}  \&
  {Heyer}}{{Heiderman} et~al.}{2010}]{heiderman:2010.gmcs}
{Heiderman} A.,  {Evans} II N.~J.,  {Allen} L.~E.,  {Huard} T.,   {Heyer} M.,
  2010, \mn@doi [\apj] {10.1088/0004-637X/723/2/1019}, \href
  {http://adsabs.harvard.edu/abs/2010ApJ...723.1019H} {723, 1019}

\bibitem[\protect\citeauthoryear{{Hennebelle} \& {Chabrier}}{{Hennebelle} \&
  {Chabrier}}{2011}]{hennebelle:2011.time.dept.imf.eps}
{Hennebelle} P.,  {Chabrier} G.,  2011, \mn@doi [\apjl]
  {10.1088/2041-8205/743/2/L29}, \href
  {http://adsabs.harvard.edu/abs/2011ApJ...743L..29H} {743, L29}

\bibitem[\protect\citeauthoryear{{Heyer}, {Gutermuth}, {Urquhart}, {Csengeri},
  {Wienen}, {Leurini}, {Menten}  \& {Wyrowski}}{{Heyer}
  et~al.}{2016}]{heyer:2016.clumps}
{Heyer} M.,  {Gutermuth} R.,  {Urquhart} J.~S.,  {Csengeri} T.,  {Wienen} M.,
  {Leurini} S.,  {Menten} K.,   {Wyrowski} F.,  2016, \mn@doi [\aap]
  {10.1051/0004-6361/201527681}, \href
  {http://adsabs.harvard.edu/abs/2016A%26A...588A..29H} {588, A29}

\bibitem[\protect\citeauthoryear{{Hills}}{{Hills}}{1980}]{hills:1980}
{Hills} J.~G.,  1980, \mn@doi [\apj] {10.1086/157703}, \href
  {http://adsabs.harvard.edu/abs/1980ApJ...235..986H} {235, 986}

\bibitem[\protect\citeauthoryear{{Hollyhead}, {Bastian}, {Adamo},
  {Silva-Villa}, {Dale}, {Ryon}  \& {Gazak}}{{Hollyhead}
  et~al.}{2015}]{hollyhead:2015.m83.ymcs}
{Hollyhead} K.,  {Bastian} N.,  {Adamo} A.,  {Silva-Villa} E.,  {Dale} J.,
  {Ryon} J.~E.,   {Gazak} Z.,  2015, \mn@doi [\mnras] {10.1093/mnras/stv331},
  \href {http://adsabs.harvard.edu/abs/2015MNRAS.449.1106H} {449, 1106}

\bibitem[\protect\citeauthoryear{{Hopkins}}{{Hopkins}}{2015}]{hopkins:gizmo}
{Hopkins} P.~F.,  2015, \mn@doi [\mnras] {10.1093/mnras/stv195}, \href
  {http://adsabs.harvard.edu/abs/2015MNRAS.450...53H} {450, 53}

\bibitem[\protect\citeauthoryear{{Hopkins} \& {Conroy}}{{Hopkins} \&
  {Conroy}}{2015}]{hopkins:2015.metal.poor}
{Hopkins} P.~F.,  {Conroy} C.,  2015, preprint, \href
  {http://adsabs.harvard.edu/abs/2015arXiv151203834H} {} (\mn@eprint {arXiv}
  {1512.03834})

\bibitem[\protect\citeauthoryear{{Hopkins} \& {Quataert}}{{Hopkins} \&
  {Quataert}}{2010}]{hopkins:zoom.sims}
{Hopkins} P.~F.,  {Quataert} E.,  2010, \mn@doi [\mnras]
  {10.1111/j.1365-2966.2010.17064.x}, \href
  {http://adsabs.harvard.edu/abs/2009arXiv0912.3257H} {407, 1529}

\bibitem[\protect\citeauthoryear{{Hopkins} \& {Raives}}{{Hopkins} \&
  {Raives}}{2016}]{hopkins:gizmo.mhd}
{Hopkins} P.~F.,  {Raives} M.~J.,  2016, \mn@doi [\mnras]
  {10.1093/mnras/stv2180}, \href
  {http://adsabs.harvard.edu/abs/2016MNRAS.455...51H} {455, 51}

\bibitem[\protect\citeauthoryear{{Hopkins}, {Murray}, {Quataert}  \&
  {Thompson}}{{Hopkins} et~al.}{2010}]{hopkins:maximum.surface.densities}
{Hopkins} P.~F.,  {Murray} N.,  {Quataert} E.,   {Thompson} T.~A.,  2010,
  \mn@doi [\mnras] {10.1111/j.1745-3933.2009.00777.x}, \href
  {http://adsabs.harvard.edu/abs/2009arXiv0908.4088H} {401, L19}

\bibitem[\protect\citeauthoryear{{Hopkins}, {Quataert}  \& {Murray}}{{Hopkins}
  et~al.}{2011}]{hopkins:rad.pressure.sf.fb}
{Hopkins} P.~F.,  {Quataert} E.,   {Murray} N.,  2011, \mn@doi [\mnras]
  {10.1111/j.1365-2966.2011.19306.x}, \href
  {http://adsabs.harvard.edu/abs/2011arXiv1101.4940H} {417, 950}

\bibitem[\protect\citeauthoryear{{Hopkins}, {Quataert}  \& {Murray}}{{Hopkins}
  et~al.}{2012a}]{hopkins:fb.ism.prop}
{Hopkins} P.~F.,  {Quataert} E.,   {Murray} N.,  2012a, \mn@doi [\mnras]
  {10.1111/j.1365-2966.2012.20578.x}, \href
  {http://adsabs.harvard.edu/abs/2012MNRAS.421.3488H} {421, 3488}

\bibitem[\protect\citeauthoryear{{Hopkins}, {Quataert}  \& {Murray}}{{Hopkins}
  et~al.}{2012b}]{hopkins:stellar.fb.winds}
{Hopkins} P.~F.,  {Quataert} E.,   {Murray} N.,  2012b, \mn@doi [\mnras]
  {10.1111/j.1365-2966.2012.20593.x}, \href
  {http://adsabs.harvard.edu/abs/2012MNRAS.421.3522H} {421, 3522}

\bibitem[\protect\citeauthoryear{{Hopkins}, {Narayanan}  \& {Murray}}{{Hopkins}
  et~al.}{2013}]{hopkins:virial.sf}
{Hopkins} P.~F.,  {Narayanan} D.,   {Murray} N.,  2013, \mn@doi [\mnras]
  {10.1093/mnras/stt723}, \href
  {http://adsabs.harvard.edu/abs/2013MNRAS.432.2647H} {432, 2647}

\bibitem[\protect\citeauthoryear{{Hopkins}, {Keres}, {Onorbe},
  {Faucher-Giguere}, {Quataert}, {Murray}  \& {Bullock}}{{Hopkins}
  et~al.}{2014}]{hopkins:2013.fire}
{Hopkins} P.~F.,  {Keres} D.,  {Onorbe} J.,  {Faucher-Giguere} C.-A.,
  {Quataert} E.,  {Murray} N.,   {Bullock} J.~S.,  2014, \mn@doi [\mnras]
  {10.1093/mnras/stu1738}, \href
  {http://adsabs.harvard.edu/abs/2013arXiv1311.2073H} {445, 581}

\bibitem[\protect\citeauthoryear{{Hopkins}, {Torrey}, {Faucher-Gigu{\`e}re},
  {Quataert}  \& {Murray}}{{Hopkins}
  et~al.}{2016}]{hopkins:qso.stellar.fb.together}
{Hopkins} P.~F.,  {Torrey} P.,  {Faucher-Gigu{\`e}re} C.-A.,  {Quataert} E.,
  {Murray} N.,  2016, \mn@doi [\mnras] {10.1093/mnras/stw289}, \href
  {http://adsabs.harvard.edu/abs/2016MNRAS.458..816H} {458, 816}

\bibitem[\protect\citeauthoryear{{Hopkins} et~al.,}{{Hopkins}
  et~al.}{2017}]{fire2}
{Hopkins} P.~F.,  et~al., 2017, preprint, \href
  {http://adsabs.harvard.edu/abs/2017arXiv170206148H} {} (\mn@eprint {arXiv}
  {1702.06148})

\bibitem[\protect\citeauthoryear{{Johnson} et~al.,}{{Johnson}
  et~al.}{2016}]{johnson:2016.cluster.formation.efficiency}
{Johnson} L.~C.,  et~al., 2016, \mn@doi [\apj] {10.3847/0004-637X/827/1/33},
  \href {http://adsabs.harvard.edu/abs/2016ApJ...827...33J} {827, 33}

\bibitem[\protect\citeauthoryear{{Kennicutt}}{{Kennicutt}}{1998a}]{kennicutt:1998.review}
{Kennicutt} Jr. R.~C.,  1998a, \mn@doi [\araa]
  {10.1146/annurev.astro.36.1.189}, \href
  {http://adsabs.harvard.edu/abs/1998ARA%26A..36..189K} {36, 189}

\bibitem[\protect\citeauthoryear{{Kennicutt}}{{Kennicutt}}{1998b}]{kennicutt98}
{Kennicutt} Jr. R.~C.,  1998b, \mn@doi [\apj] {10.1086/305588}, \href
  {http://adsabs.harvard.edu/cgi-bin/nph-bib_query?bibcode=1998ApJ...498..541K&db_key=AST}
  {498, 541}

\bibitem[\protect\citeauthoryear{{Keto}, {Ho}  \& {Lo}}{{Keto}
  et~al.}{2005}]{keto:2005.m82.gmcs}
{Keto} E.,  {Ho} L.~C.,   {Lo} K.-Y.,  2005, \mn@doi [\apj] {10.1086/497575},
  \href {http://adsabs.harvard.edu/abs/2005ApJ...635.1062K} {635, 1062}

\bibitem[\protect\citeauthoryear{{Kim} \& {Ostriker}}{{Kim} \&
  {Ostriker}}{2015}]{kim:2015.snr.mom}
{Kim} C.-G.,  {Ostriker} E.~C.,  2015, \mn@doi [\apj]
  {10.1088/0004-637X/802/2/99}, \href
  {http://adsabs.harvard.edu/abs/2015ApJ...802...99K} {802, 99}

\bibitem[\protect\citeauthoryear{{Kritsuk}, {Norman}  \& {Wagner}}{{Kritsuk}
  et~al.}{2011}]{kritsuk:2011.density.pdf.power.law}
{Kritsuk} A.~G.,  {Norman} M.~L.,   {Wagner} R.,  2011, \mn@doi [\apjl]
  {10.1088/2041-8205/727/1/L20}, \href
  {http://adsabs.harvard.edu/abs/2011ApJ...727L..20K} {727, L20}

\bibitem[\protect\citeauthoryear{{Kroupa}}{{Kroupa}}{2002}]{kroupa:imf}
{Kroupa} P.,  2002, \mn@doi [Science] {10.1126/science.1067524}, \href
  {http://adsabs.harvard.edu/abs/2002Sci...295...82K} {295, 82}

\bibitem[\protect\citeauthoryear{{Kruijssen}}{{Kruijssen}}{2012}]{kruijssen:2012.cluster.formation.efficiency}
{Kruijssen} J.~M.~D.,  2012, \mn@doi [\mnras]
  {10.1111/j.1365-2966.2012.21923.x}, \href
  {http://adsabs.harvard.edu/abs/2012MNRAS.426.3008K} {426, 3008}

\bibitem[\protect\citeauthoryear{{Kruijssen}, {Maschberger}, {Moeckel},
  {Clarke}, {Bastian}  \& {Bonnell}}{{Kruijssen}
  et~al.}{2012}]{2012MNRAS.419..841K}
{Kruijssen} J.~M.~D.,  {Maschberger} T.,  {Moeckel} N.,  {Clarke} C.~J.,
  {Bastian} N.,   {Bonnell} I.~A.,  2012, \mn@doi [\mnras]
  {10.1111/j.1365-2966.2011.19748.x}, \href
  {http://adsabs.harvard.edu/abs/2012MNRAS.419..841K} {419, 841}

\bibitem[\protect\citeauthoryear{{Krumholz}}{{Krumholz}}{2014}]{krumholz:2014.review}
{Krumholz} M.~R.,  2014, \mn@doi [\physrep] {10.1016/j.physrep.2014.02.001},
  \href {http://adsabs.harvard.edu/abs/2014PhR...539...49K} {539, 49}

\bibitem[\protect\citeauthoryear{{Krumholz} \& {Gnedin}}{{Krumholz} \&
  {Gnedin}}{2011}]{krumholz:2011.molecular.prescription}
{Krumholz} M.~R.,  {Gnedin} N.~Y.,  2011, \mn@doi [\apj]
  {10.1088/0004-637X/729/1/36}, \href
  {http://adsabs.harvard.edu/abs/2011ApJ...729...36K} {729, 36}

\bibitem[\protect\citeauthoryear{{Krumholz} \& {McKee}}{{Krumholz} \&
  {McKee}}{2005}]{krumholz.schmidt}
{Krumholz} M.~R.,  {McKee} C.~F.,  2005, \mn@doi [\apj] {10.1086/431734}, \href
  {http://adsabs.harvard.edu/abs/2005ApJ...630..250K} {630, 250}

\bibitem[\protect\citeauthoryear{{Krumholz} \& {Tan}}{{Krumholz} \&
  {Tan}}{2007}]{krumholz:sf.eff.in.clouds}
{Krumholz} M.~R.,  {Tan} J.~C.,  2007, \mn@doi [\apj] {10.1086/509101}, \href
  {http://adsabs.harvard.edu/abs/2007ApJ...654..304K} {654, 304}

\bibitem[\protect\citeauthoryear{{Krumholz} \& {Thompson}}{{Krumholz} \&
  {Thompson}}{2012}]{krumholz:2012.rrt}
{Krumholz} M.~R.,  {Thompson} T.~A.,  2012, \mn@doi [\apj]
  {10.1088/0004-637X/760/2/155}, \href
  {http://adsabs.harvard.edu/abs/2012ApJ...760..155K} {760, 155}

\bibitem[\protect\citeauthoryear{{Krumholz} \& {Thompson}}{{Krumholz} \&
  {Thompson}}{2013}]{krumholz:2013.rhd}
{Krumholz} M.~R.,  {Thompson} T.~A.,  2013, \mn@doi [\mnras]
  {10.1093/mnras/stt1174}, \href
  {http://adsabs.harvard.edu/abs/2013MNRAS.434.2329K} {434, 2329}

\bibitem[\protect\citeauthoryear{{Krumholz}, {Klein}  \& {McKee}}{{Krumholz}
  et~al.}{2011}]{krumholz:2011.rhd.starcluster.sim}
{Krumholz} M.~R.,  {Klein} R.~I.,   {McKee} C.~F.,  2011, \mn@doi [\apj]
  {10.1088/0004-637X/740/2/74}, \href
  {http://adsabs.harvard.edu/abs/2011arXiv1104.2038K} {740, 74}

\bibitem[\protect\citeauthoryear{{Krumholz}, {Dekel}  \& {McKee}}{{Krumholz}
  et~al.}{2012}]{krumholz:2012.universal.sf.efficiency}
{Krumholz} M.~R.,  {Dekel} A.,   {McKee} C.~F.,  2012, \mn@doi [\apj]
  {10.1088/0004-637X/745/1/69}, \href
  {http://adsabs.harvard.edu/abs/2012ApJ...745...69K} {745, 69}

\bibitem[\protect\citeauthoryear{{Lada} \& {Lada}}{{Lada} \&
  {Lada}}{2003}]{lada:2003.embedded.cluster.review}
{Lada} C.~J.,  {Lada} E.~A.,  2003, \mn@doi [\araa]
  {10.1146/annurev.astro.41.011802.094844}, \href
  {http://adsabs.harvard.edu/abs/2003ARA%26A..41...57L} {41, 57}

\bibitem[\protect\citeauthoryear{{Lada}, {Margulis}  \& {Dearborn}}{{Lada}
  et~al.}{1984}]{lada:1984}
{Lada} C.~J.,  {Margulis} M.,   {Dearborn} D.,  1984, \mn@doi [\apj]
  {10.1086/162485}, \href {http://adsabs.harvard.edu/abs/1984ApJ...285..141L}
  {285, 141}

\bibitem[\protect\citeauthoryear{{Larson}}{{Larson}}{1969}]{larson:1969.isothermal.collapse}
{Larson} R.~B.,  1969, \mn@doi [\mnras] {10.1093/mnras/145.3.271}, \href
  {http://adsabs.harvard.edu/abs/1969MNRAS.145..271L} {145, 271}

\bibitem[\protect\citeauthoryear{{Larson}}{{Larson}}{1981}]{larson:gmc.scalings}
{Larson} R.~B.,  1981, \mnras, \href
  {http://adsabs.harvard.edu/abs/1981MNRAS.194..809L} {194, 809}

\bibitem[\protect\citeauthoryear{{Lee}, {Chang}  \& {Murray}}{{Lee}
  et~al.}{2015}]{lee:2015.gravoturbulence}
{Lee} E.~J.,  {Chang} P.,   {Murray} N.,  2015, \mn@doi [\apj]
  {10.1088/0004-637X/800/1/49}, \href
  {http://adsabs.harvard.edu/abs/2015ApJ...800...49L} {800, 49}

\bibitem[\protect\citeauthoryear{{Lee}, {Miville-Desch{\^e}nes}  \&
  {Murray}}{{Lee} et~al.}{2016}]{lee:2016.gmc.eff}
{Lee} E.~J.,  {Miville-Desch{\^e}nes} M.-A.,   {Murray} N.~W.,  2016, \mn@doi
  [\apj] {10.3847/1538-4357/833/2/229}, \href
  {http://adsabs.harvard.edu/abs/2016ApJ...833..229L} {833, 229}

\bibitem[\protect\citeauthoryear{{Martizzi}, {Faucher-Gigu{\`e}re}  \&
  {Quataert}}{{Martizzi} et~al.}{2015}]{martizzi:2015.snr.inhomogeneous}
{Martizzi} D.,  {Faucher-Gigu{\`e}re} C.-A.,   {Quataert} E.,  2015, \mn@doi
  [\mnras] {10.1093/mnras/stv562}, \href
  {http://adsabs.harvard.edu/abs/2015MNRAS.450..504M} {450, 504}

\bibitem[\protect\citeauthoryear{{Mathieu}}{{Mathieu}}{1983}]{mathieu:1983}
{Mathieu} R.~D.,  1983, \mn@doi [\apjl] {10.1086/184011}, \href
  {http://adsabs.harvard.edu/abs/1983ApJ...267L..97M} {267, L97}

\bibitem[\protect\citeauthoryear{{Mayer}, {Kazantzidis}, {Escala}  \&
  {Callegari}}{{Mayer} et~al.}{2010}]{mayer:2009.direct.collapse.bh}
{Mayer} L.,  {Kazantzidis} S.,  {Escala} A.,   {Callegari} S.,  2010, \mn@doi
  [\nat] {10.1038/nature09294}, \href
  {http://adsabs.harvard.edu/abs/2010Natur.466.1082M} {466, 1082}

\bibitem[\protect\citeauthoryear{{Mayer}, {Fiacconi}, {Bonoli}, {Quinn}, {Ro{\v
  s}kar}, {Shen}  \& {Wadsley}}{{Mayer}
  et~al.}{2015}]{mayer:2015.direct.collapse.bh}
{Mayer} L.,  {Fiacconi} D.,  {Bonoli} S.,  {Quinn} T.,  {Ro{\v s}kar} R.,
  {Shen} S.,   {Wadsley} J.,  2015, \mn@doi [\apj]
  {10.1088/0004-637X/810/1/51}, \href
  {http://adsabs.harvard.edu/abs/2015ApJ...810...51M} {810, 51}

\bibitem[\protect\citeauthoryear{{McCrady} \& {Graham}}{{McCrady} \&
  {Graham}}{2007}]{mccrady:m82.sscs}
{McCrady} N.,  {Graham} J.~R.,  2007, \mn@doi [\apj] {10.1086/518357}, \href
  {http://adsabs.harvard.edu/abs/2007ApJ...663..844M} {663, 844}

\bibitem[\protect\citeauthoryear{{Mouri} \& {Taniguchi}}{{Mouri} \&
  {Taniguchi}}{2002}]{mouri:2002.runaway.black.holes}
{Mouri} H.,  {Taniguchi} Y.,  2002, \mn@doi [\apjl] {10.1086/339472}, \href
  {http://adsabs.harvard.edu/abs/2002ApJ...566L..17M} {566, L17}

\bibitem[\protect\citeauthoryear{{Murray}}{{Murray}}{2011}]{murray:2010.sfe.mw.gmc}
{Murray} N.,  2011, \mn@doi [\apj] {10.1088/0004-637X/729/2/133}, \href
  {http://adsabs.harvard.edu/abs/2010arXiv1007.3270M} {729, 133}

\bibitem[\protect\citeauthoryear{{Murray}, {Quataert}  \& {Thompson}}{{Murray}
  et~al.}{2010}]{murray:molcloud.disrupt.by.rad.pressure}
{Murray} N.,  {Quataert} E.,   {Thompson} T.~A.,  2010, \mn@doi [\apj]
  {10.1088/0004-637X/709/1/191}, \href
  {http://adsabs.harvard.edu/abs/2009arXiv0906.5358M} {709, 191}

\bibitem[\protect\citeauthoryear{{Myers}, {Klein}, {Krumholz}  \&
  {McKee}}{{Myers} et~al.}{2014}]{2014MNRAS.439.3420M}
{Myers} A.~T.,  {Klein} R.~I.,  {Krumholz} M.~R.,   {McKee} C.~F.,  2014,
  \mn@doi [\mnras] {10.1093/mnras/stu190}, \href
  {http://adsabs.harvard.edu/abs/2014MNRAS.439.3420M} {439, 3420}

\bibitem[\protect\citeauthoryear{{Nordlund} \& {Padoan}}{{Nordlund} \&
  {Padoan}}{1999}]{padoan:1999.density.pdf}
{Nordlund} {\AA}.~K.,  {Padoan} P.,  1999, in {Franco} J.,  {Carraminana} A.,
  eds, Interstellar Turbulence. p.~218 (\mn@eprint {} {astro-ph/9810074})

\bibitem[\protect\citeauthoryear{{Orr} et~al.,}{{Orr}
  et~al.}{2017}]{orr:2016.what.fires.up.SF}
{Orr} M.,  et~al., 2017, preprint, \href
  {http://adsabs.harvard.edu/abs/2017arXiv170101788O} {} (\mn@eprint {arXiv}
  {1701.01788})

\bibitem[\protect\citeauthoryear{{Ostriker} \& {Shetty}}{{Ostriker} \&
  {Shetty}}{2011}]{ostriker.shetty:2011}
{Ostriker} E.~C.,  {Shetty} R.,  2011, \mn@doi [\apj]
  {10.1088/0004-637X/731/1/41}, \href
  {http://adsabs.harvard.edu/abs/2011ApJ...731...41O} {731, 41}

\bibitem[\protect\citeauthoryear{{Padoan}, {Nordlund}  \& {Jones}}{{Padoan}
  et~al.}{1997}]{padoan:1997.density.pdf}
{Padoan} P.,  {Nordlund} A.,   {Jones} B.~J.~T.,  1997, \mnras, \href
  {http://adsabs.harvard.edu/abs/1997MNRAS.288..145P} {288, 145}

\bibitem[\protect\citeauthoryear{{Padoan}, {Haugb{\o}lle}  \&
  {Nordlund}}{{Padoan} et~al.}{2012}]{padoan:2012.sfe}
{Padoan} P.,  {Haugb{\o}lle} T.,   {Nordlund} {\AA}.,  2012, \mn@doi [\apjl]
  {10.1088/2041-8205/759/2/L27}, \href
  {http://adsabs.harvard.edu/abs/2012ApJ...759L..27P} {759, L27}

\bibitem[\protect\citeauthoryear{{Parra}, {Conway}, {Diamond}, {Thrall},
  {Lonsdale}, {Lonsdale}  \& {Smith}}{{Parra} et~al.}{2007}]{arp.220.fast.sf2}
{Parra} R.,  {Conway} J.~E.,  {Diamond} P.~J.,  {Thrall} H.,  {Lonsdale} C.~J.,
   {Lonsdale} C.~J.,   {Smith} H.~E.,  2007, \mn@doi [\apj] {10.1086/511813},
  \href {http://adsabs.harvard.edu/abs/2007ApJ...659..314P} {659, 314}

\bibitem[\protect\citeauthoryear{{Penston}}{{Penston}}{1969}]{penston:1969.isothermal.collapse}
{Penston} M.~V.,  1969, \mn@doi [\mnras] {10.1093/mnras/144.4.425}, \href
  {http://adsabs.harvard.edu/abs/1969MNRAS.144..425P} {144, 425}

\bibitem[\protect\citeauthoryear{{Portegies Zwart} \& {McMillan}}{{Portegies
  Zwart} \& {McMillan}}{2002}]{portegies-zwart:2002}
{Portegies Zwart} S.~F.,  {McMillan} S.~L.~W.,  2002, \mn@doi [\apj]
  {10.1086/341798}, \href {http://adsabs.harvard.edu/abs/2002ApJ...576..899P}
  {576, 899}

\bibitem[\protect\citeauthoryear{{Portegies Zwart}, {McMillan}  \&
  {Gieles}}{{Portegies Zwart}
  et~al.}{2010}]{portegies-zwart:2010.starcluster.review}
{Portegies Zwart} S.~F.,  {McMillan} S.~L.~W.,   {Gieles} M.,  2010, \mn@doi
  [\araa] {10.1146/annurev-astro-081309-130834}, \href
  {http://adsabs.harvard.edu/abs/2010ARA%26A..48..431P} {48, 431}

\bibitem[\protect\citeauthoryear{{Rafikov}}{{Rafikov}}{2007}]{rafikov:2007.convect.cooling.grav.instab.planets}
{Rafikov} R.~R.,  2007, \mn@doi [\apj] {10.1086/517599}, \href
  {http://adsabs.harvard.edu/abs/2007ApJ...662..642R} {662, 642}

\bibitem[\protect\citeauthoryear{{Raskutti}, {Ostriker}  \&
  {Skinner}}{{Raskutti} et~al.}{2016}]{raskutti:2016.gmcs}
{Raskutti} S.,  {Ostriker} E.~C.,   {Skinner} M.~A.,  2016, preprint, \href
  {http://adsabs.harvard.edu/abs/2016arXiv160804469R} {} (\mn@eprint {arXiv}
  {1608.04469})

\bibitem[\protect\citeauthoryear{{Rees}}{{Rees}}{1976}]{rees:1976.opacity.limit}
{Rees} M.~J.,  1976, \mn@doi [\mnras] {10.1093/mnras/176.3.483}, \href
  {http://adsabs.harvard.edu/abs/1976MNRAS.176..483R} {176, 483}

\bibitem[\protect\citeauthoryear{{Richings}, {Schaye}  \&
  {Oppenheimer}}{{Richings} et~al.}{2014a}]{richlings:2014a}
{Richings} A.~J.,  {Schaye} J.,   {Oppenheimer} B.~D.,  2014a, \mn@doi [\mnras]
  {10.1093/mnras/stu525}, \href
  {http://adsabs.harvard.edu/abs/2014MNRAS.440.3349R} {440, 3349}

\bibitem[\protect\citeauthoryear{{Richings}, {Schaye}  \&
  {Oppenheimer}}{{Richings} et~al.}{2014b}]{richlings:2014b}
{Richings} A.~J.,  {Schaye} J.,   {Oppenheimer} B.~D.,  2014b, \mn@doi [\mnras]
  {10.1093/mnras/stu1046}, \href
  {http://adsabs.harvard.edu/abs/2014MNRAS.442.2780R} {442, 2780}

\bibitem[\protect\citeauthoryear{{Rodriguez}, {Morscher}, {Pattabiraman},
  {Chatterjee}, {Haster}  \& {Rasio}}{{Rodriguez}
  et~al.}{2015}]{rodriguez:2015.bbh.globulars}
{Rodriguez} C.~L.,  {Morscher} M.,  {Pattabiraman} B.,  {Chatterjee} S.,
  {Haster} C.-J.,   {Rasio} F.~A.,  2015, \mn@doi [Physical Review Letters]
  {10.1103/PhysRevLett.115.051101}, \href
  {http://adsabs.harvard.edu/abs/2015PhRvL.115e1101R} {115, 051101}

\bibitem[\protect\citeauthoryear{{Rodriguez}, {Chatterjee}  \&
  {Rasio}}{{Rodriguez} et~al.}{2016}]{rodriguez:2016.bbh.globulars}
{Rodriguez} C.~L.,  {Chatterjee} S.,   {Rasio} F.~A.,  2016, \mn@doi [\prd]
  {10.1103/PhysRevD.93.084029}, \href
  {http://adsabs.harvard.edu/abs/2016PhRvD..93h4029R} {93, 084029}

\bibitem[\protect\citeauthoryear{{Rosdahl} \& {Teyssier}}{{Rosdahl} \&
  {Teyssier}}{2015}]{rosdahl:2015.rhd}
{Rosdahl} J.,  {Teyssier} R.,  2015, \mn@doi [\mnras] {10.1093/mnras/stv567},
  \href {http://adsabs.harvard.edu/abs/2015MNRAS.449.4380R} {449, 4380}

\bibitem[\protect\citeauthoryear{{Ryon} et~al.,}{{Ryon}
  et~al.}{2015}]{ryon:2015.m83.clusters}
{Ryon} J.~E.,  et~al., 2015, \mn@doi [\mnras] {10.1093/mnras/stv1282}, \href
  {http://adsabs.harvard.edu/abs/2015MNRAS.452..525R} {452, 525}

\bibitem[\protect\citeauthoryear{{Saitoh}, {Daisaka}, {Kokubo}, {Makino},
  {Okamoto}, {Tomisaka}, {Wada}  \& {Yoshida}}{{Saitoh}
  et~al.}{2008}]{saitoh:2008.highres.disks.high.sf.thold}
{Saitoh} T.~R.,  {Daisaka} H.,  {Kokubo} E.,  {Makino} J.,  {Okamoto} T.,
  {Tomisaka} K.,  {Wada} K.,   {Yoshida} N.,  2008, \pasj, \href
  {http://adsabs.harvard.edu/abs/2008PASJ...60..667S} {60, 667}

\bibitem[\protect\citeauthoryear{{Scoville} et~al.,}{{Scoville}
  et~al.}{2017}]{scoville:2016.arp220}
{Scoville} N.,  et~al., 2017, \mn@doi [\apj] {10.3847/1538-4357/836/1/66},
  \href {http://adsabs.harvard.edu/abs/2017ApJ...836...66S} {836, 66}

\bibitem[\protect\citeauthoryear{{Skinner} \& {Ostriker}}{{Skinner} \&
  {Ostriker}}{2013}]{skinner:2013.hyperion}
{Skinner} M.~A.,  {Ostriker} E.~C.,  2013, \mn@doi [\apjs]
  {10.1088/0067-0049/206/2/21}, \href
  {http://adsabs.harvard.edu/abs/2013ApJS..206...21S} {206, 21}

\bibitem[\protect\citeauthoryear{{Skinner} \& {Ostriker}}{{Skinner} \&
  {Ostriker}}{2015}]{skinner:2015.ir.molcloud.disrupt}
{Skinner} M.~A.,  {Ostriker} E.~C.,  2015, \mn@doi [\apj]
  {10.1088/0004-637X/809/2/187}, \href
  {http://adsabs.harvard.edu/abs/2015ApJ...809..187S} {809, 187}

\bibitem[\protect\citeauthoryear{{Smith} \& {Gallagher}}{{Smith} \&
  {Gallagher}}{2001}]{smith:2001.m82.ssc}
{Smith} L.~J.,  {Gallagher} J.~S.,  2001, \mn@doi [\mnras]
  {10.1046/j.1365-8711.2001.04627.x}, \href
  {http://adsabs.harvard.edu/abs/2001MNRAS.326.1027S} {326, 1027}

\bibitem[\protect\citeauthoryear{{Smith}, {Slater}, {Fellhauer}, {Goodwin}  \&
  {Assmann}}{{Smith} et~al.}{2011}]{smith:2011.cluster.assembly}
{Smith} R.,  {Slater} R.,  {Fellhauer} M.,  {Goodwin} S.,   {Assmann} P.,
  2011, \mn@doi [\mnras] {10.1111/j.1365-2966.2011.19039.x}, \href
  {http://adsabs.harvard.edu/abs/2011MNRAS.416..383S} {416, 383}

\bibitem[\protect\citeauthoryear{{Smith}, {Goodwin}, {Fellhauer}  \&
  {Assmann}}{{Smith} et~al.}{2013}]{smith:2013.cluster.assembly}
{Smith} R.,  {Goodwin} S.,  {Fellhauer} M.,   {Assmann} P.,  2013, \mn@doi
  [\mnras] {10.1093/mnras/sts106}, \href
  {http://adsabs.harvard.edu/abs/2013MNRAS.428.1303S} {428, 1303}

\bibitem[\protect\citeauthoryear{{Solomon}, {Rivolo}, {Barrett}  \&
  {Yahil}}{{Solomon} et~al.}{1987}]{solomon:gmc.scalings}
{Solomon} P.~M.,  {Rivolo} A.~R.,  {Barrett} J.,   {Yahil} A.,  1987, \mn@doi
  [\apj] {10.1086/165493}, \href
  {http://adsabs.harvard.edu/abs/1987ApJ...319..730S} {319, 730}

\bibitem[\protect\citeauthoryear{{Springel}}{{Springel}}{2005}]{springel:gadget}
{Springel} V.,  2005, \mn@doi [\mnras] {10.1111/j.1365-2966.2005.09655.x},
  \href
  {http://adsabs.harvard.edu/cgi-bin/nph-bib_query?bibcode=2005MNRAS.364.1105S&db_key=AST}
  {364, 1105}

\bibitem[\protect\citeauthoryear{Springel}{Springel}{2010}]{springel:arepo}
Springel V.,  2010, \mn@doi [\mnras] {10.1111/j.1365-2966.2009.15715.x}, \href
  {http://adsabs.harvard.edu/abs/2010MNRAS.401..791S} {401, 791}

\bibitem[\protect\citeauthoryear{{Springel} \& {Hernquist}}{{Springel} \&
  {Hernquist}}{2003}]{springel:multiphase}
{Springel} V.,  {Hernquist} L.,  2003, \mn@doi [\mnras]
  {10.1046/j.1365-8711.2003.06206.x}, \href
  {http://adsabs.harvard.edu/cgi-bin/nph-bib_query?bibcode=2003MNRAS.339..289S&db_key=AST}
  {339, 289}

\bibitem[\protect\citeauthoryear{{Stone}, {Gardiner}, {Teuben}, {Hawley}  \&
  {Simon}}{{Stone} et~al.}{2008}]{stone:2008.athena}
{Stone} J.~M.,  {Gardiner} T.~A.,  {Teuben} P.,  {Hawley} J.~F.,   {Simon}
  J.~B.,  2008, \mn@doi [\apjs] {10.1086/588755}, \href
  {http://adsabs.harvard.edu/abs/2008ApJS..178..137S} {178, 137}

\bibitem[\protect\citeauthoryear{{Su}, {Hopkins}, {Hayward}, {Faucher-Giguere},
  {Keres}, {Ma}  \& {Robles}}{{Su} et~al.}{2016}]{su:2016.feedback.first}
{Su} K.-Y.,  {Hopkins} P.~F.,  {Hayward} C.~C.,  {Faucher-Giguere} C.-A.,
  {Keres} D.,  {Ma} X.,   {Robles} V.~H.,  2016, preprint, \href
  {http://adsabs.harvard.edu/abs/2016arXiv160705274S} {} (\mn@eprint {arXiv}
  {1607.05274})

\bibitem[\protect\citeauthoryear{{Tan}, {Krumholz}  \& {McKee}}{{Tan}
  et~al.}{2006}]{tan:2006}
{Tan} J.~C.,  {Krumholz} M.~R.,   {McKee} C.~F.,  2006, \mn@doi [\apjl]
  {10.1086/504150}, \href {http://adsabs.harvard.edu/abs/2006ApJ...641L.121T}
  {641, L121}

\bibitem[\protect\citeauthoryear{{Thompson} \& {Krumholz}}{{Thompson} \&
  {Krumholz}}{2016}]{thompson:2016.eddington.outflows}
{Thompson} T.~A.,  {Krumholz} M.~R.,  2016, \mn@doi [\mnras]
  {10.1093/mnras/stv2331}, \href
  {http://adsabs.harvard.edu/abs/2016MNRAS.455..334T} {455, 334}

\bibitem[\protect\citeauthoryear{{Thompson}, {Quataert}  \&
  {Murray}}{{Thompson} et~al.}{2005}]{thompson:rad.pressure}
{Thompson} T.~A.,  {Quataert} E.,   {Murray} N.,  2005, \mn@doi [\apj]
  {10.1086/431923}, \href {http://adsabs.harvard.edu/abs/2005ApJ...630..167T}
  {630, 167}

\bibitem[\protect\citeauthoryear{{Torrey}, {Hopkins}, {Faucher-Gigu{\`e}re},
  {Vogelsberger}, {Quataert}, {Kere{\v s}}  \& {Murray}}{{Torrey}
  et~al.}{2016}]{torrey:2016.feedback.instability}
{Torrey} P.,  {Hopkins} P.~F.,  {Faucher-Gigu{\`e}re} C.-A.,  {Vogelsberger}
  M.,  {Quataert} E.,  {Kere{\v s}} D.,   {Murray} N.,  2016, preprint, \href
  {http://adsabs.harvard.edu/abs/2016arXiv160107186T} {} (\mn@eprint {arXiv}
  {1601.07186})

\bibitem[\protect\citeauthoryear{{Troland} \& {Crutcher}}{{Troland} \&
  {Crutcher}}{2008}]{troland:2012.gmc.zeeman}
{Troland} T.~H.,  {Crutcher} R.~M.,  2008, \mn@doi [\apj] {10.1086/587546},
  \href {http://adsabs.harvard.edu/abs/2008ApJ...680..457T} {680, 457}

\bibitem[\protect\citeauthoryear{{Tsang} \& {Milosavljevi{\'c}}}{{Tsang} \&
  {Milosavljevi{\'c}}}{2015}]{tsang:2015.rhd}
{Tsang} B.~T.-H.,  {Milosavljevi{\'c}} M.,  2015, \mn@doi [\mnras]
  {10.1093/mnras/stv1707}, \href
  {http://adsabs.harvard.edu/abs/2015MNRAS.453.1108T} {453, 1108}

\bibitem[\protect\citeauthoryear{{Tsz-Ho Tsang} \& {Milosavljevic}}{{Tsz-Ho
  Tsang} \& {Milosavljevic}}{2017}]{tsang:2017.ssc.rp}
{Tsz-Ho Tsang} B.,  {Milosavljevic} M.,  2017, preprint, \href
  {http://adsabs.harvard.edu/abs/2017arXiv170907539T} {} (\mn@eprint {arXiv}
  {1709.07539})

\bibitem[\protect\citeauthoryear{{Tutukov}}{{Tutukov}}{1978}]{tutukov:1978}
{Tutukov} A.~V.,  1978, \aap, \href
  {http://adsabs.harvard.edu/abs/1978A%26A....70...57T} {70, 57}

\bibitem[\protect\citeauthoryear{{Utreras}, {Becerra}  \& {Escala}}{{Utreras}
  et~al.}{2016}]{utreras:2016.rotation}
{Utreras} J.,  {Becerra} F.,   {Escala} A.,  2016, \mn@doi [\apj]
  {10.3847/0004-637X/833/1/13}, \href
  {http://adsabs.harvard.edu/abs/2016ApJ...833...13U} {833, 13}

\bibitem[\protect\citeauthoryear{{Vazquez-Semadeni}}{{Vazquez-Semadeni}}{1994}]{vazquez-semadeni:1994.turb.density.pdf}
{Vazquez-Semadeni} E.,  1994, \mn@doi [\apj] {10.1086/173847}, \href
  {http://adsabs.harvard.edu/abs/1994ApJ...423..681V} {423, 681}

\bibitem[\protect\citeauthoryear{{Vink}, {de Koter}  \& {Lamers}}{{Vink}
  et~al.}{2001}]{vink:2001.ob.mass.loss}
{Vink} J.~S.,  {de Koter} A.,   {Lamers} H.~J.~G.~L.~M.,  2001, \mn@doi [\aap]
  {10.1051/0004-6361:20010127}, \href
  {http://adsabs.harvard.edu/abs/2001A%26A...369..574V} {369, 574}

\bibitem[\protect\citeauthoryear{{Vutisalchavakul}, {Evans}  \&
  {Heyer}}{{Vutisalchavakul} et~al.}{2016}]{vuti:2016.gmcs}
{Vutisalchavakul} N.,  {Evans} II N.~J.,   {Heyer} M.,  2016, \mn@doi [\apj]
  {10.3847/0004-637X/831/1/73}, \href
  {http://adsabs.harvard.edu/abs/2016ApJ...831...73V} {831, 73}

\bibitem[\protect\citeauthoryear{{Wilson}, {Harris}, {Longden}  \&
  {Scoville}}{{Wilson} et~al.}{2006}]{wilson:2006.arp220.superclumps}
{Wilson} C.~D.,  {Harris} W.~E.,  {Longden} R.,   {Scoville} N.~Z.,  2006,
  \mn@doi [\apj] {10.1086/500577}, \href
  {http://adsabs.harvard.edu/abs/2006ApJ...641..763W} {641, 763}

\bibitem[\protect\citeauthoryear{{Wu}, {Evans}, {Gao}, {Solomon}, {Shirley}  \&
  {Vanden Bout}}{{Wu} et~al.}{2005}]{wu:2005.clumps}
{Wu} J.,  {Evans} II N.~J.,  {Gao} Y.,  {Solomon} P.~M.,  {Shirley} Y.~L.,
  {Vanden Bout} P.~A.,  2005, \mn@doi [\apjl] {10.1086/499623}, \href
  {http://adsabs.harvard.edu/abs/2005ApJ...635L.173W} {635, L173}

\bibitem[\protect\citeauthoryear{{Wu}, {Evans}, {Shirley}  \& {Knez}}{{Wu}
  et~al.}{2010}]{wu:2010.clumps}
{Wu} J.,  {Evans} II N.~J.,  {Shirley} Y.~L.,   {Knez} C.,  2010, \mn@doi
  [\apjs] {10.1088/0067-0049/188/2/313}, \href
  {http://adsabs.harvard.edu/abs/2010ApJS..188..313W} {188, 313}

\bibitem[\protect\citeauthoryear{{Yoast-Hull}, {Gallagher}  \&
  {Zweibel}}{{Yoast-Hull} et~al.}{2016}]{starburst.cosmic.rays}
{Yoast-Hull} T.~M.,  {Gallagher} J.~S.,   {Zweibel} E.~G.,  2016, \mn@doi
  [\mnras] {10.1093/mnrasl/slv195}, \href
  {http://adsabs.harvard.edu/abs/2016MNRAS.457L..29Y} {457, L29}

\bibitem[\protect\citeauthoryear{{Zhang} \& {Davis}}{{Zhang} \&
  {Davis}}{2017}]{zhang:2017.ir.pressure}
{Zhang} D.,  {Davis} S.~W.,  2017, \mn@doi [\apj] {10.3847/1538-4357/aa6935},
  \href {http://adsabs.harvard.edu/abs/2017ApJ...839...54Z} {839, 54}

\makeatother
\end{thebibliography}

\begin{appendix}

\section{Code Tests} 
\begin{figure}
\includegraphics[width=\columnwidth]{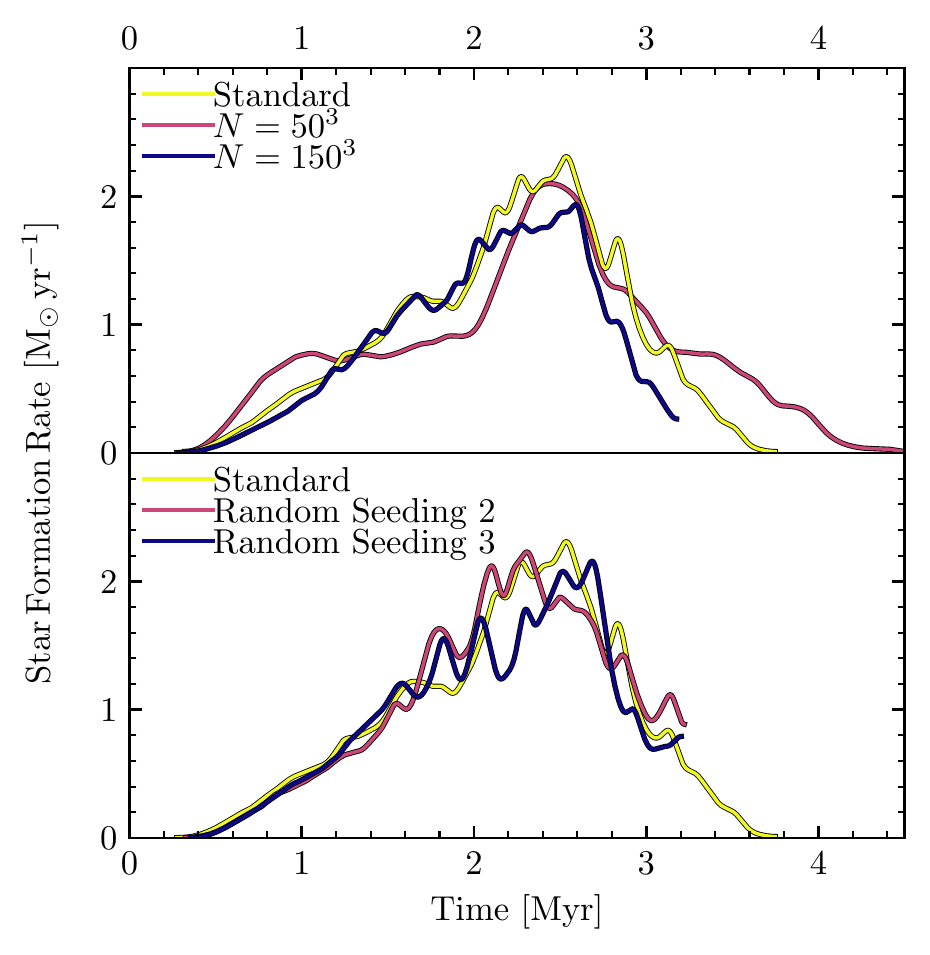}
\caption{Star formation histories of test runs with parameters $M=\unit[10^7]{\msun}$ and $R=\unit[50]{pc}$. {\it Top:} Convergence tests with particle number varied from $50^3$ to $200^3$. {\it Bottom:} Consistency tests using 3 different random seeds for the initial perturbations.}
\label{SFRConvergence}
\end{figure}
\subsection{Convergence and consistency}
\label{convergence}
The methods for cooling, star formation and feedback used in this paper have been tested in previous studies of galactic-scale simulations resolving spatial scales of $\sim \unit[1]{pc}$ and masses $>\unit[10^3]{\msun}$. However, their behaviour at the higher resolutions of these simulations has been much less well-studied. It is therefore necessary to determine how the simulation behaviour depends (1) upon mass and spatial resolution, (2) upon the particular random seeding in the initial conditions and (3) upon the particular physics included and parameters chosen. Because the star formation histories (SFH) are the main data of interest, we shall focus on the effects of these choices on the SFH as a proxy for the behaviour of the simulation as a whole.

We choose the parameters $R=\unit[50]{pc}$, $M=\unit[10^7]{\msun}$ as the point in parameter space at which to investigate these questions. Because all runs are qualitatively identical with only differences in numerical scalings, the conclusions drawn for these parameters should apply across our parameter space, obviating the need to perform the tests at all points. We vary the particle number from $50^3$ to $150^3$ to isolate resolution effects. Because we use adaptive softening, the effective gravitational force resolution naturally follows mass resolution with no need for manual tuning. To assess the effect of the random velocity seeding, we compare runs from 3 random realizations at the standard resolution and with standard physics. 

From the first panel of Figure \ref{SFRConvergence} it is evident that mass resolution does have certain systematic effects upon the computed SFH: in particular, low-resolution runs have a SFR which is greater at early times. This is an artifact the cutoff in the turbulent length scale that can be followed before the turbulent Jeans mass is no longer resolved. A gas structure that is well-resolved and supported against its self-gravity by internal motions at high resolution may not be considered so if down-sampled to low resolution where it consists only of a few particles. Thus, in the absence of any feedback moderation, as is the case at early times, the SFR will rise sooner at low resolution. While this resolution effect is conspicuous, it apparently does not have a strong effect upon the integrated SFE.

The variation in SFE due to resolution is in fact comparable to the variation arising from different random seedings at fixed resolution, visible in panel 2 of Figure \ref{CompareSFR}. In both cases, the mass of gas converted to stars varies only by$\sim1\%$ between runs. We therefore conclude that the star formation efficiencies computed as the central result of this study are consistent between runs with the same physical parameters. As discussed in the main text, our results concerning star formation efficiency can be understood in terms of simple force balance considerations. As such, it is not surprising that the SFE should converge rapidly and be robust with respect to perturbations.

\subsection{Radiation pressure} 
\label{appendix:radtest}
\begin{figure}
\includegraphics[width=\columnwidth]{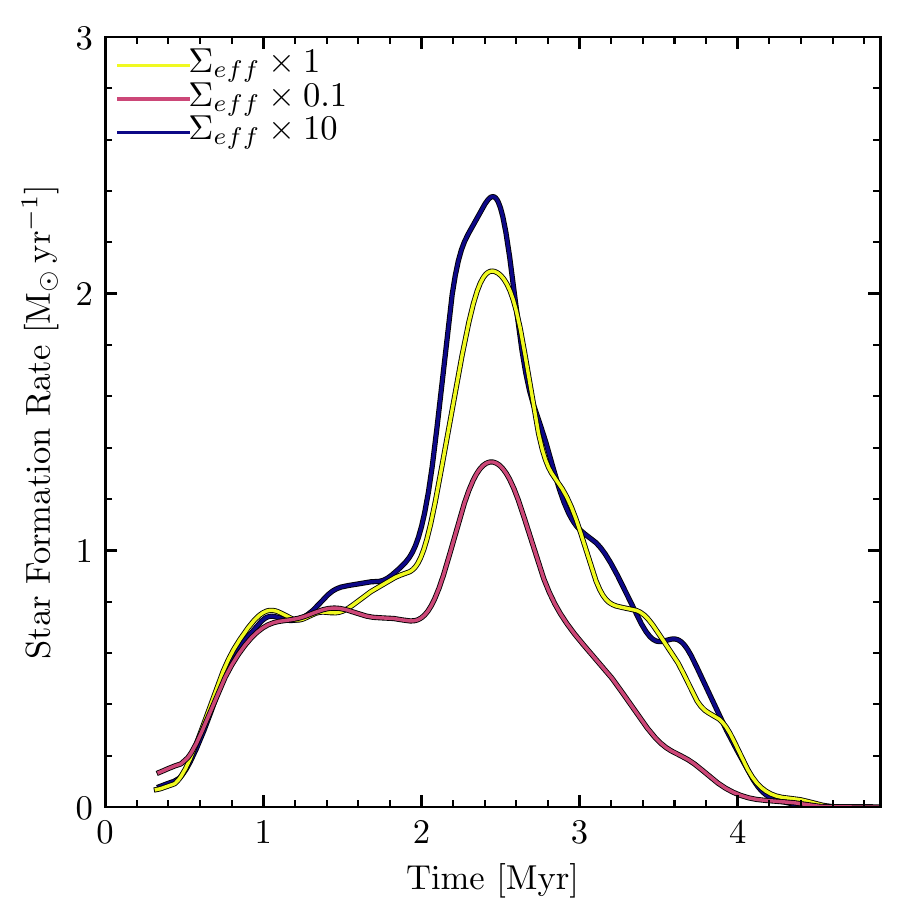}
\caption{Effect of varying the local extinction column density estimator $\Sigma_{eff}$ by factors of 0.1 and 10 in our treatment of radiation pressure.}
\label{fig:radtest}
\end{figure}
\label{rptest}
In our survey of the effects of different physics (Section \ref{physicssurvey}), it was found that radiation pressure was most responsible for the moderation of star formation. Therefore, it is particularly important to test the robustness of the radiative transfer prescription we have used. Radiation pressure is treated with a combination of short-ranged, local coupling within the kernel encompassing a star particle's nearest neighbouring gas particles (most importantly handing single-scattered UV photons), and a long-ranged component treated in the optically-thin approximation (mainly handling reprocessed IR photons). The estimated local extinction around a particle relies upon an estimate of the local column density $\Sigma_{eff}$ obtained by a Sobolev approximation; for details see \citet{fire2}. 

To test the sensitivity of our results to this local extinction approximation, we both increased and decreased the estimated $\Sigma_{eff}$ by a factor of 10 in our fiducial $10^7\msun$, $50\mathrm{pc}$ run at $50^3$ resolution. The resulting star formation histories are shown in Figure \ref{fig:radtest}. Increasing $\Sigma_{eff}$ by a factor of 10 had very little effect on the the star formation history. This is because the local extinction fraction $1-\exp\left(\Sigma_{eff}\kappa_{UV}\right)$ is typically already quite close to 1 in the default run. Decreasing $\Sigma_{eff}$ by a factor of 10 reduced the peak SFR by roughly a factor of 2, and decreased the final SFE from 0.32 to 0.23, as leakage of UV photons from the local kernel is increased. We conclude that the SFE results of this paper do have some amount of sensitivity to the assumed geometric factor in the prescription for $\Sigma_{eff}$, but this sensitivity is quite sublinear: variations of a factor of 10 lead to SFE variations within a factor of 2.

Finally, we also performed a series of radiation pressure-only tests with a cloud of mass $\unit[5\times10^4]{\msun}$ and radius $\unit[15]{pc}$, with a statistically-isotropic solenoidal initial turbulent velocity field scaled to give an initial virial parameter of $2$, emulating the setup in \citet{raskutti:2016.gmcs}. At mass resolutions at which the formation of dense protostellar envelopes starts to be resolved ($<<\unit[1]{\msun}$), one might worry that some qualitative change in the nature of the density field would affect the local column density estimates in such a way that the net photon momentum budget at large is affected, and hence the SFE. We ran this test with particle numbers of $12^3$, $25^3$, $50^3$, and $100^3$, and obtained $\epsilon_{int}$ of $0.082$, $0.052$, $0.042$, and $0.040$ respectively, suggesting convergence. As with our convergence test with all physics enabled (\ref{convergence}), the SFE tends to converge from above; the star formation criterion is in some sense stricter at higher resolution, as local velocity gradients supporting against gravitational collapse are better-resolved.

\end{appendix}

\end{document}